\pdfoutput=1
\documentclass[12pt]{elsarticle}
\usepackage{blindtext}
\usepackage{amsmath}
\usepackage[usenames, dvipsnames]{xcolor}
\usepackage[utf8x]{inputenc}
\usepackage[T1]{fontenc}
\usepackage{tikz}
\usepackage{pgfplots, pgfplotstable}
\usetikzlibrary{calc,arrows,fadings,decorations.pathreplacing,decorations.markings,patterns,shapes.geometric}
\usepackage{fp}
\usepackage{yhmath}
\usepackage{verbatim}
\usepackage{enumitem}
\usepackage{tikz}

\usepackage{color,soul}
\usepackage{amssymb}
\usepackage{caption}
\usepackage{subcaption}
\usetikzlibrary{calc,fit,intersections,shapes}
\usetikzlibrary{calc}
\usepgfplotslibrary{polar}
\pgfplotsset{compat=1.6}
\pgfplotsset{every axis plot}
\journal{ }
\usepackage{natbib}
\usepackage{amsmath}
\usepackage[normalem]{ulem}
\usepackage{multirow}

\begin{document}

\title{Simultaneous Energy Harvesting and Bearing Fault Detection using Piezoelectric Cantilevers}
\author[label1]{P. Peralta-Braz}
\author[label1]{M. M. Alamdari\footnote{Corresponding author, m.makkialamdari@unsw.edu.au}}
\author[label2]{C. T. Chou}
\author[label2]{M. Hassan}
\author[label1]{E. Atroshchenko}

\address[label1]{School of Civil and Environmental Engineering, University of New South Wales, Sydney, Australia}
\address[label2]{School of Computer Science and Engineering, University of New South Wales, Sydney, Australia}

\journal{IEEE Internet of Things Journal}
\begin{abstract}
Bearings are critical components in industrial machinery, yet their vulnerability to faults often leads to costly breakdowns. Conventional fault detection methods depend on continuous, high-frequency vibration sensing, digitizing, and wireless transmission to the cloud—an approach that significantly drains the limited energy reserves of battery-powered sensors, accelerating their depletion and increasing maintenance costs. This work proposes a fundamentally different approach: rather than using instantaneous vibration data, we employ piezoelectric energy harvesters (PEHs) tuned to specific frequencies and leverage the cumulative harvested energy over time as the key diagnostic feature. By directly utilizing the energy generated from the machinery’s vibrations, we eliminate the need for frequent analog-to-digital conversions and data transmission, thereby reducing energy consumption at the sensor node and extending its operational lifetime. To validate this approach, we use a numerical PEH model and publicly available acceleration datasets, examining various PEH designs with different natural frequencies. We also consider the influence of the classification algorithm, the number of devices, and the observation window duration. The results demonstrate that the harvested energy reliably indicates bearing faults across a range of conditions and severities. By converting vibration energy into both a power source and a diagnostic feature, our solution offers a more sustainable, low-maintenance strategy for fault detection in smart machinery.
\end{abstract}

\maketitle


\section{Introduction}
\label{S:1}
In modern industry, bearings are indispensable components, particularly in rotating machinery such as compressors, engines, and motors. Their high load-carrying capacity makes them essential in critical sectors like rail transportation, wind turbines, and aerospace \cite{he2023real}. However, their operation in harsh environments makes rolling bearings susceptible to failure, accounting for 45–55\% of all mechanical breakdowns \cite{jin2022time}. Recent reports indicate that the world’s largest manufacturers collectively lose approximately \$1 trillion annually due to machine failures \cite{senseye2022downtime}.  Bearing failure is a progressive process that advances through distinct stages: normal operation, incipient failure, severe failure, and ultimate failure \cite{rai2018integrated}. Therefore, early detection and timely intervention are crucial for maintaining operational continuity and optimizing industrial efficiency.

Fault detection methods can be classified into two main categories: model-based and data-driven approaches \cite{xu2020intelligent}. Model-driven approaches rely on prior physical knowledge; however, precisely characterizing the internal stress and other physical quantities in running rolling bearings presents a significant challenge \cite{shi2020research}. This limitation can hinder the generalization of models and their capacity to represent the complex, nonlinear nature of bearing degradation \cite{wen2024new}. For this reason, data-driven approaches have gained significant attention as a more suitable strategy for failure detection in large-scale industrial processes. These approaches infer the bearing's health condition by analysing data collected from operating machinery \cite{mushtaq2021deep}. Common data sources include acoustic signals \cite{pacheco2022bearing}, motor current signals \cite{zhang2022motor,guan2024enhancing}, and vibration signals \cite{tiboni2022review}. Among these, vibration signals are particularly valuable due to their sensitivity to abnormal rotor equipment conditions, making it possible to identify inherent fault frequencies \cite{he2023real}. Recently, several vibration-based methods utilising deep learning \cite{hakim2023systematic, zhang2020deep}, and machine learning \cite{soomro2024insights, zhang2021machine} have been introduced. These methods generally follow three steps \cite{saini2022predictive}: (1) data pre-processing, (2) feature extraction, and (3) pattern recognition. Cloud computing \cite{bahga2011analyzing} is often employed to perform fault detection tasks. Thus, data is collected and transmitted to a centralized cloud server, where it is pre-processed, features are extracted, and the condition is evaluated.

Overall, various data-driven approaches discussed in the literature demonstrate promising performance. However, advancing efforts to integrate these methods with cost-effective hardware solutions that simultaneously minimise disruptions to rotating systems remains crucial \cite{fu2023edgecog}. In this context, smart bearings have recently emerged as a promising concept. These advanced systems integrate bearing structures with sensors, signal processing, and wireless communication to enable real-time condition monitoring and fault detection in rotating machinery. This allows for failure prediction, optimised maintenance schedules, and enhanced system reliability by collecting and analysing data such as vibration, temperature, load, and speed \cite{gong2022variable,zhang2023comprehensive}. Additionally, recent advancements in energy harvesting (EH) and energy storage technologies, along with decreasing costs, have further accelerated the development of self-powered Internet of Things (IoT) devices \cite{shirvanimoghaddam2019towards}, such as smart bearing.  Since vibrations act as both an information and energy source, Piezoelectric Energy Harvesters (PEHs) are particularly effective for powering smart bearing systems. They offer key advantages such as high energy efficiency per unit area, simple geometry, and scalability.

\begin{figure}[h!]
	\centering
    \includegraphics[width=1.0\textwidth]{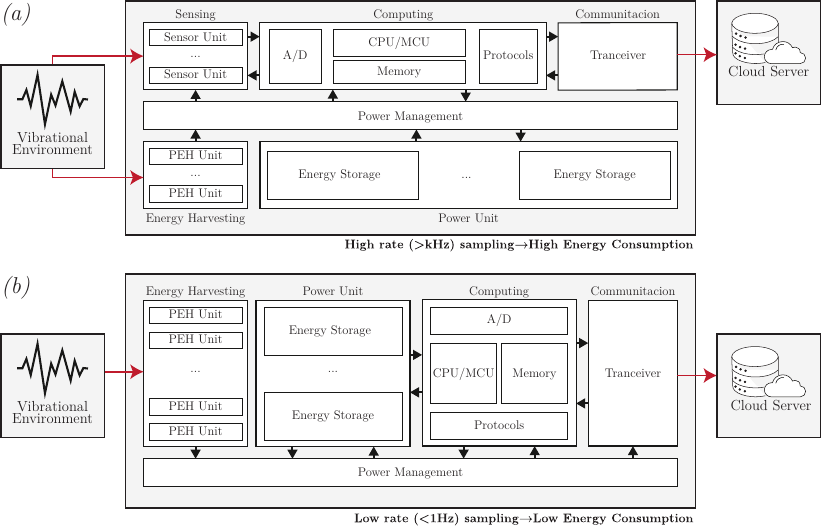}
\caption{$(a)$ A standard setup of a self-powered wireless IoT device. $(b)$ Proposed setup of a self-powered wireless IoT device}
	\label{Ch5:fig:intro}
\end{figure}

The current standard setup of a self-powered IoT device comprises an energy harvester, sensors, a processing unit, a power management module, energy storage, and a transceiver \cite{shirvanimoghaddam2019towards}, as shown in Figure \ref{Ch5:fig:intro}$a$. The PEH plays a critical role by converting mechanical energy into electrical energy. However, predicting the amount of harvested energy is challenging due to the stochastic nature of the vibration source \cite{shaikh2016energy}. This unpredictable behaviour makes it impractical to directly power the device, necessitating the inclusion of a power management module \cite{adu2018energy}. The power management module coordinates power distribution among the device's units to balance power generation with consumption. Its primary objective is to achieve energy-neutral operation (ENO) \cite{sharma2010optimal}. On the other hand, the energy storage system typically consists of a combination of rechargeable batteries and supercapacitors, designed to meet the power requirements of the devices \cite{shirvanimoghaddam2019towards}. Meanwhile, sensors are employed to collect accurate signal data. To achieve self-powered operation, minimizing sensor power consumption is crucial. For instance, accelerometers typically consume power in the milliwatt range \cite{zhang2023comprehensive}. The computing module digitalizes the sensor signals by analog-to-digital (A/D) converters and the sampled data are then
transferred wirelessly to a cloud server for further processing. The transceiver is responsible for both transmitting signal data to the cloud and receiving control messages from terminals to the smart bearing \cite{zhang2023comprehensive}. Selecting an appropriate transceiver technology is a crucial aspect of designing self-powered devices. Accurate analysis of a machine's condition often requires data to be collected and transmitted at high sampling rates, typically in the tens of kilohertz range \cite{das2017industrial}. However, this requirement presents challenges as it significantly increases the energy consumption of wireless transmission due to the higher volume of data. Additionally, the energy usage of A/D conversion rises significantly under these conditions. The overall energy consumption for this process of data collection is approximately proportional to the sampling frequency \cite{mulleti2023power}. Although significant efforts have been made in the literature to reduce the energy consumption of data collection, primarily through minimizing energy use in wireless transmissions \cite{tanash2023enhancing} or by duty cycling \cite{chu2023iris}, there appears to be little focus on addressing the energy consumption bottleneck caused by high-frequency sampling.

Therefore, achieving a viable self-powered system requires minimising the power consumption associated with $(i)$ sensing and $(ii)$ data collection. A promising concept introduced in the literature to address the first challenge, sensing power consumption, is the Simultaneous Energy Harvesting and Sensing (SEHS) system \cite{ma2020simultaneous}. This approach enables a piezoelectric energy harvester (PEH) to serve a dual purpose, working both as an energy harvester and as a sensor to extract valuable information to infer the operating conditions. On the other hand, to reduce the power consumption associated with data collection, edge computing \cite{qian2019edge} has emerged as a powerful solution. By processing and analysing data at the network's edge rather than in the cloud, it minimizes the volume of data sent to remote servers. This approach ensures efficient handling of critical data while managing less essential information locally. Furthermore, by incorporating analogue computation \cite{zangeneh2021analogue}, the features used for fault classification are computed directly in the analogue system before the signal passes through the A/D module, ensuring simplified and efficient processing and avoiding the energy costs associated with wireless transmission of high-frequency samples.

A new fault detection architecture that significantly reduces energy consumption by utilising PEHs is proposed, as shown in Figure \ref{Ch5:fig:intro}$b$. In this approach, we remove dedicated sensors, and delegate sensing task to a bank of PEHs, which are specifically designed to be tuned to inherent frequencies. Our method involves measuring the amount of energy captured periodically as a feature for the fault classification task. The concept is based on the premise that the voltage signal generated from bearing vibrations inherently contains critical operational information. As a result, the harvested energy of a PEH can be directly used for fault diagnosis. By tuning each PEH to a key frequency, it becomes capable of detecting the presence or absence of specific harmonics in the signal, effectively identifying degradation and spall formation in the bearing. This architecture provides a promising solution for performing data pre-processing and feature extraction at the network edge, thereby reducing the need for extensive digitisation and subsequently minimising the data transmitted to a cloud server for pattern recognition tasks.
In this work, we explore the potential of using the energy generated by PEHs as a feature for fault detection in bearings. The study evaluates ten different PEH designs, utilising a numerical model \cite{peralta2020parametric} to estimate the harvested energy, and tests them on publicly available datasets, the Case Western Reserve University (CWRU) dataset \cite{zhang2021machine} and the Intelligent Maintenance Systems (IMS) dataset \cite{sacerdoti2023comparison}. The use of public datasets is driven by the objective of establishing a benchmark scenario widely recognized within the research community. This choice not only enhances comparability with prior studies but also ensures a standardized and robust validation of the results. Additionally, we investigate the influence of factors such as geometry, the number of devices, time window length, and the electrical circuit on fault detection performance. As a next step, we propose a methodology for online incipient fault detection, which is subsequently tested on diverse public datasets.

The contributions of this paper can be summarised as follows:
\begin{itemize}
    \item Present a decision-making framework for fault detection in bearings that significantly reduces the amount of collected data, thereby decreasing energy consumption in digitization and transmission.
    \item Provide evidence, for the first time, that the amount of capacitor energy harvested in by PEHs within a pre-defined time window can be effectively used as a feature for bearing fault classification.
    \item Conduct a rigorous study to understand the impact of piezoelectric devices' resonance frequency, number of devices, classification algorithms, and time windows on fault classification using a public bearing database.

\end{itemize}

The remainder of the paper is organised as follows. Related work is reviewed in Section \ref{Ch5:S:RW}. Section \ref{Ch5:S:2} presents the PEH numerical model used to estimate the generated energy from bearing accelerations. Section \ref{Ch5:S:3} describes the harvested energy database obtained from the CWRU acceleration database, which is utilised in subsequent sections. Section \ref{Ch5:S:4} addresses the damage classification problem using the harvested energy dataset. The following sections tackle practical considerations: Section \ref{Ch5:S:5} examines the impact of the time window length, while Section \ref{Ch5:S:6} examines the use and influence of a capacitor within an electrical circuit to estimate the harvested energy. Section \ref{Ch5:S:7} focuses on the anomaly detection problem to target online monitoring applications. Finally, conclusions are provided in Section \ref{Ch5:S:8}.

\newpage
\section{Related Work}
\label{Ch5:S:RW}
The use of piezoelectric devices designed to harvest energy as sensors has been explored in the literature across a range of applications, including infrastructure \cite{cahill2018vibration, fitzgerald2019scour, liu2021energy, krishnanunni2023efficacy}, human recognition \cite{ma2020simultaneous, safaei2018energy, choudhry2020flexible, gogoi2024choice, umetsu2019ehaas}, and machinery \cite{ rashidi2018dual, de2021design, wang2023energy, zhang2022piezoelectric, shi2024research}. In this sense, Zhang et al. \cite{zhang2022piezoelectric} investigated this concept for bearing fault diagnosis by installing an arc-shaped piezoelectric sheet between the outer race of a rolling bearing and the bearing pedestal. The study shows that the output voltage signal can identify frequencies rich in information related to the bearing's fault condition. Specifically, the study considers rolling bearings with inner and outer race defects and, by analysing the signals, recognizes fault characteristic frequency components in their spectrums. This finding motivates further investigation, particularly as the presence or absence of fault characteristic frequencies in the voltage signal of the piezoelectric device influences the amount of harvested energy. This effect is especially pronounced if the device is designed as a monostable, resonance frequency-tuned system. By tuning the device’s resonance to a key frequency, the harvested energy can serve as a classification feature.


However, Safian et al. \cite{ safian2023development} identified some issues in the system proposed by Zhang et al. The first issue concerns the mounting of the PZT sheet, which alters the bearing housing. This alteration could affect the bearing's performance, including its rotational speed, load capacity, and lifespan. The second issue relates to the brittleness of piezoelectric ceramics, which can be subjected to high stresses under shaft radial loads. This limitation confines the design to low-load machines. The third issue concerns the system's sensitivity, particularly given that the studies were conducted in a noise- and vibration-free laboratory environment. 

Notably, Shi et al. \cite{shi2024research}  address some of the aforementioned issues by proposing a system based on a monostable piezoelectric device. This system offers the advantage of simple installation without requiring significant modifications to the bearing's structure and avoiding excessive stress on the piezoelectric material. In this work, the electrical output signal is analysed using a CNN-based framework for fault classification, demonstrating excellent performance with accuracies exceeding 98\%. While this work demonstrates the potential of using a monostable piezoelectric device tuned to a fault characteristic frequency, its primary focus is not on minimizing the energy consumption associated with data digitization and transmission. Notably, the sampling frequency used is 50 kHz, meaning the system generates 50,000 voltage samples per second. This high data rate poses challenges for both digitization, which requires substantial processing power, and wireless transmission, which demands significant energy—factors that can be critical in energy-constrained applications.



In this context, Lan et al. \cite{lan2017capsense} explored the use of accumulated energy in a capacitor for activity sensing, demonstrating strong performance. While their study focused on human activity, their approach holds significant potential for bearing applications. This is particularly relevant given the variations in characteristic frequencies between states and the generally stationary nature of these machines over time. A capacitor-based sensing approach enables decision-making based on a \textit{single} measurement of the capacitor’s energy after a set period. Compared to the methodology proposed by Shi et al.\cite{shi2024research}, which collects each sample over 20 seconds at a 50 kHz sampling rate, this approach dramatically reduces the amount of data that needs to be digitized and transmitted—from 20 million bytes to just 2 bytes (assuming each sample requires 2 bytes for storage.).

\section{Piezoelectric Energy Harvester Model}
\label{Ch5:S:2}
\begin{figure}[h]
	\centering
	\includegraphics[width=0.9\textwidth]{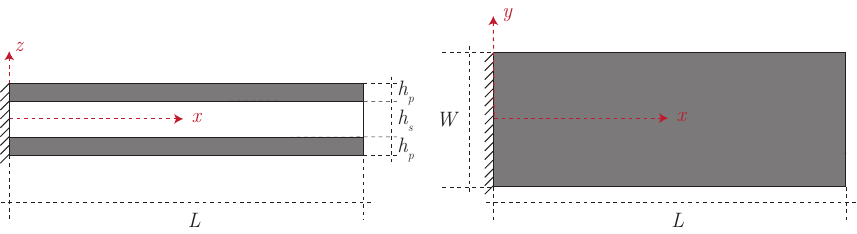}
\caption{Illustration of the piezoelectric energy harvester, designed as a cantilever plate, with two piezoelectric layers and one substructure layer.}
	\label{Ch5:fig:piezodevice}
\end{figure}

To estimate the energy harvested by a PEH, this study utilises a model based on Kirchhoff-Love plate theory and Hamilton's principle for electro-mechanical systems. It is solved numerically using the IsoGeometric Analysis (IGA) approach. The model presented in \cite{peralta2020parametric} demonstrated significant accuracy at a reasonable computational expense. An illustration of a PEH is presented in Figure \ref{Ch5:fig:piezodevice}. The device is considered to have a rectangular structure with width $W$ and length $L$, consisting of a substructure layer of thickness $h_s$ and two piezoelectric layers of thickness $h_p$, attached to a vibrating source and connected to an electrical resistance.

The IGA utilizes B-Splines $N_I$ to parameterize the device's domain and approximate the relative deflection of the mid-plane with a vector of control variables $\textbf{w}\in\mathbb{R}^{N \times 1}$, where $N$ denotes the total number of degrees of freedom. Additionally, a Modal Order Reduction method is utilised, which involves approximating $\textbf{w}$ using a truncated expansion of the first $K$ mode shape vectors, expressed as $\textbf{w} \approx \textbf{w}_o = \boldsymbol{\Phi}_o \boldsymbol{\eta}$. In this context, $\boldsymbol{\Phi}_o\in\mathbb{R}^{N \times K}$ represents the matrix including the first $K$ mode shape vectors $\boldsymbol{\phi}_i$, whereas $\boldsymbol{\eta}\in\mathbb{R}^{K\times1}$ signifies the modal coordinates. Consequently, the approach results in a coupled system of differential equations represented as
\begin{equation}
\label{Ch5:SistEq1R}
\textbf{I}_o\ddot{\boldsymbol{\eta}}+\textbf{c}_o\dot{\boldsymbol{\eta}}+ \textbf{k}_o\boldsymbol{\eta} - \boldsymbol{\theta}_o v(t) = \textbf{f}_o a_b(t)
\end{equation}

\begin{equation}
\label{Ch5:SistEq2R}
    C_p\dot{v}(t) + \frac{v(t)}{R_l} + \boldsymbol{\Theta}^T\boldsymbol{\Phi}_o \dot{\boldsymbol{\eta}}=0 
\end{equation}
where  the equation (\ref{Ch5:SistEq1R}) represents the reduced mechanical equation of motion with electrical coupling, while the equation (\ref{Ch5:SistEq2R}) denotes the reduced electrical circuit equation with mechanical coupling. $\mathbf{I}_o\in\mathbb{R}^{K\times K}$ is the identity matrix. $\mathbf{k}_o\in\mathbb{R}^{K\times K}$ is the reduced stiffness matrix, $\mathbf{c}_o\in\mathbb{R}^{K\times K}$ is the reduced mechanical damping matrix, $\mathbf{f}_o\in\mathbb{R}^{K\times1}$ is the mechanical forces vector, $\boldsymbol{\Theta}\in\mathbb{R}^{N\times1}$ is the reduced electro-mechanical coupling vector, $\boldsymbol{\theta}_o\in\mathbb{R}^{K\times1}$ is the electro-mechanical coupling vector; $C_p$ is the capacitance and $R_l$ is the external electric resistance; $a_b(t)$ is the base acceleration and $v(t)$ is the output voltage. For more details refer to \cite{peralta2023design}.

By performing time integration of the system (\ref{Ch5:SistEq1R})-(\ref{Ch5:SistEq2R}), the voltage signal induced by an arbitrary base acceleration $a_b(t)$ is obtained. In this work, $a_b(t)$ represents the acceleration signal from a specific bearing operational state. The Matlab \verb!ode45! solver is employed to solve this system for $v(t)$. Furthermore, the associated electrical energy ($E$) produced during the time interval $t \in [t_1, t_2]$, can be determined using the following equation:
\begin{equation}
    E = \int_{t_1}^{t_2} \frac{v^2(t)}{R_l} dt \label{Ch5:eqEnergy} 
\end{equation}

In a specific scenario where the base acceleration is a harmonic signal, represented as $a_b(t) = A_b e^{i \omega t}$ (where $i=\sqrt{-1}$), it can be demonstrated that the output voltage is also harmonic, expressed as $v(t) = V_o e^{i\omega t}$. The Frequency Response Function (FRF), denoted as $H = H(\omega)$, can be define to correlate the amplitudes of the output voltage $V_o$ and the excitation acceleration $A_b$ for a given frequency $\omega$, based on equations (\ref{Ch5:SistEq1R}) and (\ref{Ch5:SistEq2R}): 
\small
\begin{equation}
    \label{Ch5:Hv_equation}
    H(\omega) = i\omega\left(\frac{1}{R_l}+i\omega C_p\right)^{-1} \boldsymbol{\Theta}^T\boldsymbol{\Phi}_o\left(-\omega^2\textbf{I}_o+j\omega \textbf{c}_o + \textbf{k}_o + i\omega \left(\frac{1}{R_l}+i\omega C_p\right)^{-1} \boldsymbol{\theta}_o \boldsymbol{\Theta}^T\boldsymbol{\Phi}_o\right)^{-1}\textbf{f}_o
\end{equation}
\normalsize

\section{Case Western Reserve University (CWRU) database}
\label{Ch5:S:3}
\begin{figure}[h!]
	\centering
	\includegraphics[width=0.9\textwidth]{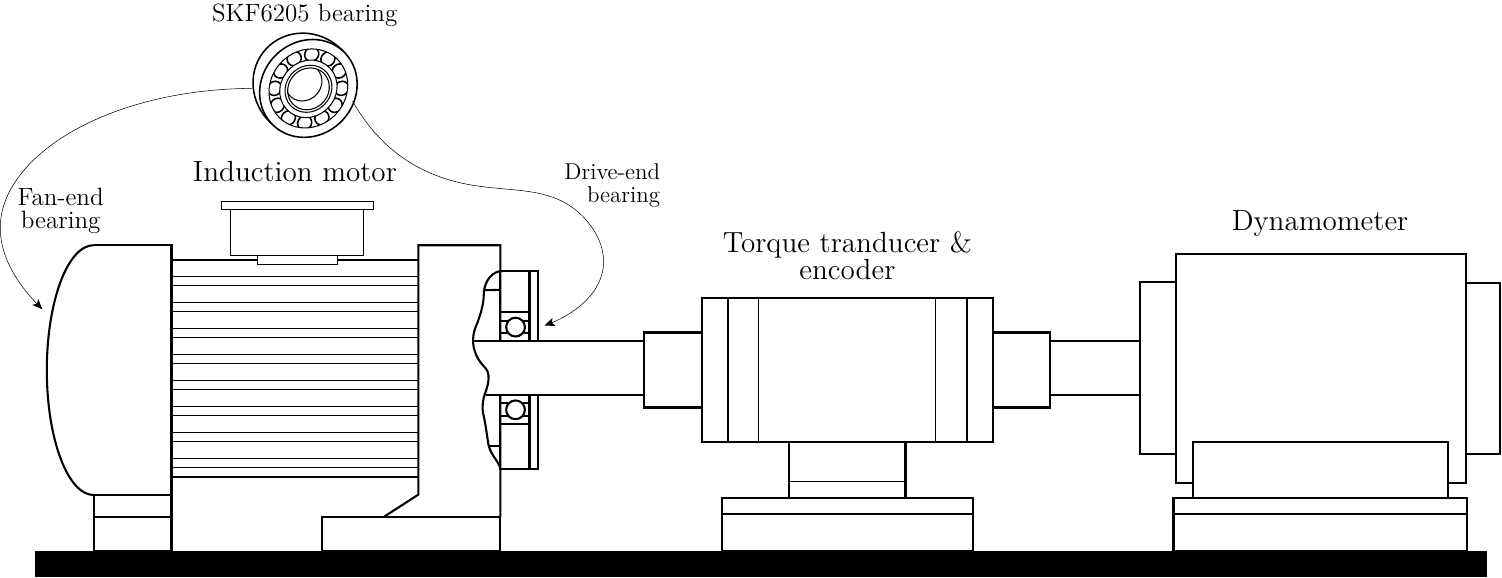}
\caption{Bearing experimental setup of CWRU database}
	\label{Ch5:fig:setup_CWRU}
\end{figure}

The Case Western Reserve University (CWRU) Bearing Center is the most extensively utilised public database for the validation of bearing fault classification methods. Figure \ref{Ch5:fig:setup_CWRU} illustrates the test bench of the CWRU database, comprising a 2 hp electric motor, a torque sensor, and a power dynamometer. Accelerometers are affixed to the casings of both the drive end and the fan end to collect vibration data. Torque is applied on the shaft by a dynamometer and an electronic control system.

The dataset comprises vibration data from an electric motor with either healthy or artificially damaged roller bearings at the drive-end or fan-end. Data was collected using two accelerometers: one mounted vertically on the drive-end bearing housing and the other on the fan-end bearing housing. The drive-end bearing data was recorded at either 12,000 or 48,000 Hz, while the fan-end bearing data was consistently recorded at 12,000 Hz. However, the baseline or normal condition was recorded at 12,000 Hz, so only data at this frequency are considered for this study.

This dataset includes four distinct states: normal condition, inner race faults (IRF), ball faults (BF), and outer race faults (ORF). ORCs are further classified based on the fault location at the 6 o’clock, 3 o’clock, and 12 o’clock positions. The fault diameters range from 0.007 inches to 0.028 inches. Electro-discharge machining introduced these faults, causing artificial damage to the bearings. SKF bearings were used for the smaller diameters (0.007, 0.014, and 0.021 inches), while NTN equivalent bearings were used for the largest diameter (0.028 inches).

However, not all combinations of damage conditions, fault locations, and fault severities are fully reported. To have consistency in terms of damage conditions and severity, only a subset of the data will be considered in the present work. For instance, we exclude all datasets with a fault diameter of 0.028 inches from the analysis, as the dataset for ORFs is not available. Additionally, for a fault diameter of 0.014 inches at the outer race fault, it is only reported at the 6 o'clock position. Therefore, moving forward, when referring to ORFs, we will exclusively consider the 6 o'clock position.

Therefore, this study examines ten conditions that include the normal state and various combinations of damage types (BF, ORF, and IRF) with different fault diameters. The fault diameters are denominated based on severity: slight (0.007 inches), medium (0.014 inches), and severe (0.021 inches). For simplicity, we focus exclusively on the bearings at the drive end, operating at a constant speed under a motor load of 1 hp.

\begin{figure}[h!]
	\centering
	\includegraphics[width=1.0\textwidth]{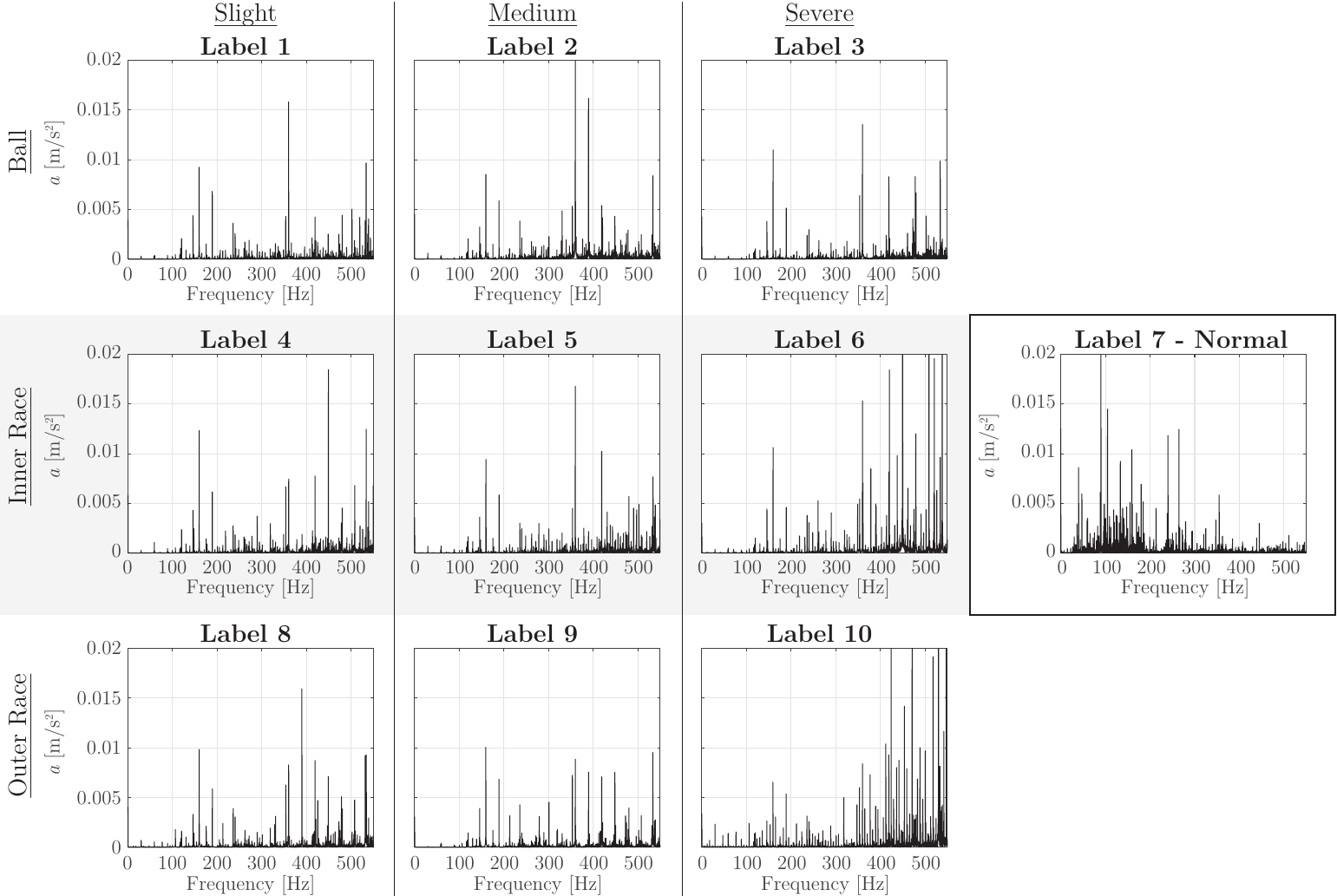}
\caption{FFT of the ten labels, considering the normal condition and the combination of fault location (ball, inner race, and outer race) and fault severity (slight, medium, and severe).}
	\label{Ch5:fig:10FFT}
\end{figure}

In this study, we define ten distinct structural conditions, each represented by a specific label. These include: \textbf{Label 1} for Slight BF, \textbf{Label 2} for Medium BF, \textbf{Label 3} for Severe BF, \textbf{Label 4} for Slight IRF, \textbf{Label 5} for Medium IRF, \textbf{Label 6} for Severe IRF, \textbf{Label 7} for Normal, \textbf{Label 8} for Slight ORF, \textbf{Label 9} for Medium ORF, and \textbf{Label 10} for Severe ORF. For each label, a 10-second signal is analysed, despite the availability of a 40-second signal for the normal condition. This decision ensures consistency across all conditions by using equal-length signals. Figure \ref{Ch5:fig:10FFT} presents the Fast Fourier Transform (FFT) of the acceleration signals for the ten labels. Its range is truncated to 0–500 Hz, which is the normal operating range for piezoelectric energy harvesters (PEH). The figure shows distinct characteristics in the frequency domain for each condition, highlighting variations that could impact the energy output of a PEH.

The impact of the PEH design on the damage identification problem is analysed by studying ten different devices. These devices are composed of PZT5A attached to a steel substructure. The devices differ by their length $L$, which aims to achieve a natural frequency within the range of 50-500 Hz with a step of 50 Hz, while all other geometrical parameters are kept identical. Specifically, the thickness of the piezoelectric layers $h_p$ and substructure $h_s$ are 1 mm, and the width $W$ is 50 mm. Table \ref{Ch5:t:Landwo} presents the length ($L$) and natural frequency ($\omega_0$) of the ten piezoelectric devices.

\begin{figure}[h]
	\centering
    \includegraphics[width=1.0\textwidth]{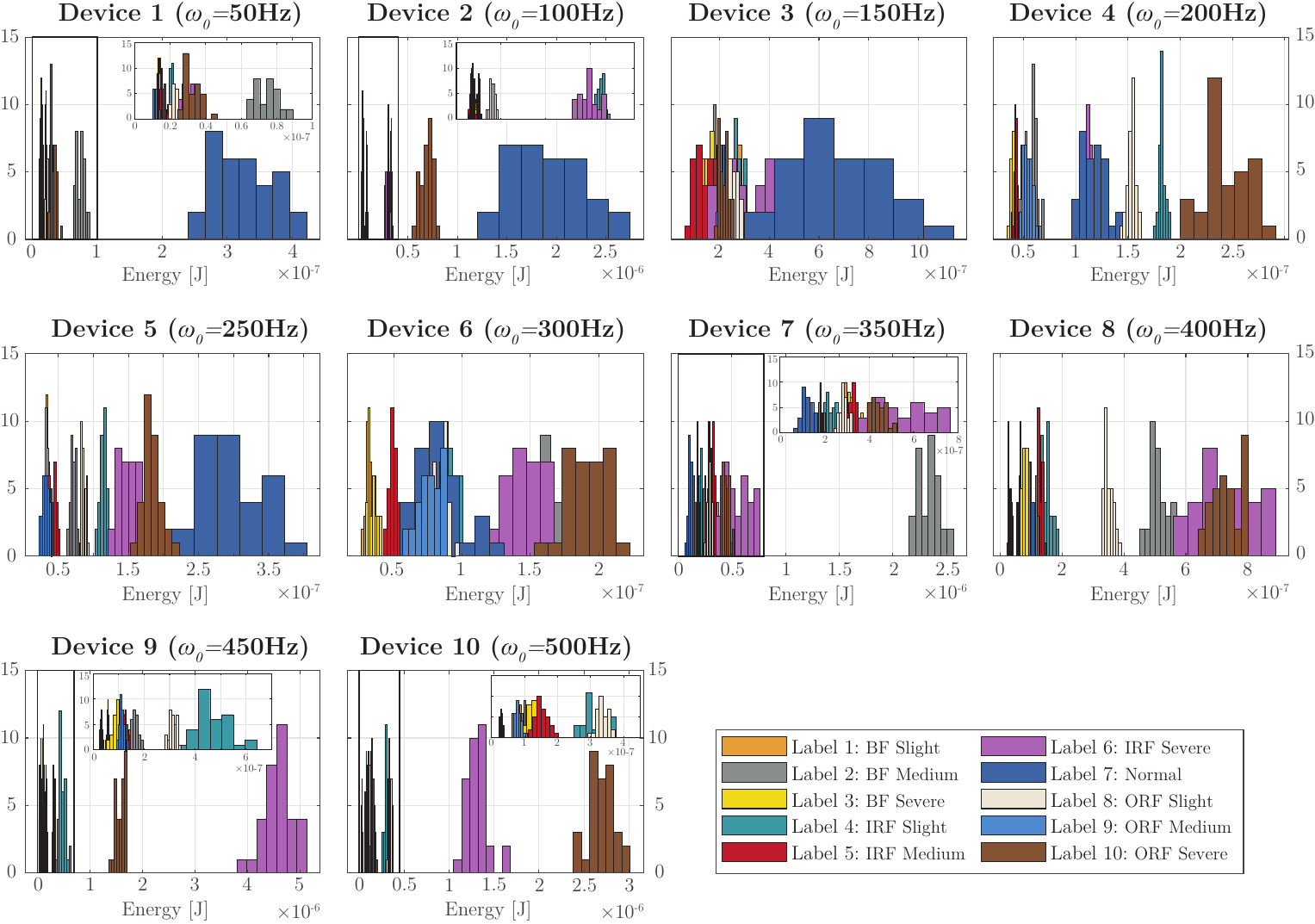}
\caption{Histograms of the energy harvested by each of the ten devices under the ten labelled fault conditions}
	\label{Ch5:fig:EnergyHist}
\end{figure}

\begin{table}[h]
\centering
\caption{Lengths ($L$) and natural frequencies ($\omega_0$) of the ten piezoelectric devices used in the study.}
\renewcommand{\arraystretch}{1.25}
\label{Ch5:t:Landwo}
\begin{tabular}{ccrrrrrrrrr}
\hline
\multicolumn{1}{l}{}         & \multicolumn{10}{c}{Device}                                                                                                                                                                                                                        \\ \cline{2-11} 
                             & 1                         & \multicolumn{1}{c}{2} & \multicolumn{1}{c}{3} & \multicolumn{1}{c}{4} & \multicolumn{1}{c}{5} & \multicolumn{1}{c}{6} & \multicolumn{1}{c}{7} & \multicolumn{1}{c}{8} & \multicolumn{1}{c}{9} & \multicolumn{1}{c}{10} \\ \hline
$\omega_0$ (Hz)              & \multicolumn{1}{r}{50}    & 100                   & 150                   & 200                   & 250                   & 300                   & 350                   & 400                   & 450                   & 500                    \\
\multicolumn{1}{r}{$L$ (cm)} & \multicolumn{1}{r}{14.10} & 10.25                 & 8.40                  & 7.28                  & 6.52                  & 5.97                  & 5.52                  & 5.17                  & 4.88                  & 4.63                   \\ \hline
\end{tabular}
\end{table}
Section \ref{Ch5:S:2} presents ten acceleration signals, each corresponding to one of the 10 bearing condition labels, with a duration of 10 seconds. Each acceleration signal is divided into 33 events (or non-overlapping signals) lasting 0.3 seconds (3600 sampling points). The energy harvested by the 10 devices is then calculated using Eq. \ref{Ch5:eqEnergy}. The harvested energy from the $events$ under each labeled condition, using each of the PEHs, is represented in histograms. This analysis illustrates the magnitude, variability, and dispersion of the harvested energy. Figure \ref{Ch5:fig:EnergyHist} displays the results, where each plot shows the histogram of energy harvested by a single device across the 10 condition labels.

As expected, for each device, the average amount of harvested energy corresponding to a specific state varies between labels. However, the energy from $events$ under the same label shows variability, which can cause overlap between the values of different labels, potentially complicating classification when using harvested energy as a feature. Additionally, the histograms across devices reveal a lack of correlation, suggesting that some devices may perform better in classification than others. While this provides initial insight, it is too early to draw definitive conclusions. We resume this discussion in the next section, where we explore the classification problem in greater detail.


\section{Bearing Fault Classification based on the Harvested Energy}
\label{Ch5:S:4}
In this section, a series of classification methods are tested on the database of harvested energy from the ten piezoelectric devices described in Section \ref{Ch5:S:2}. The main objective is to present an initial approach that demonstrates the potential of harvested energy as a valuable feature for classifying the ten labels, which include normal conditions and various damage states with different locations and severity. Since this work does not focus on innovating classification methods, some of the most well-known techniques are selected for this study, including: (a) K-Nearest Neighbors (KNN), (b) Support Vector Machine (SVM), (c) Random Forest, and (d) Naive Bayes. A brief description of each method is presented next.
\begin{enumerate}[label=\alph*)]
    \item KNN is one of the simplest and most intuitive methods in data classification. The KNN algorithm's fundamental concept is that it can classify a sample into a category if it shares similarities with the nearest $K$ samples in the feature space \cite{bishop2006pattern, cover1967nearest}.
    \item SVM is one of the most widely used and effective classification algorithms. It constructs an optimal hyperplane in the feature space that maximises the margin between different classes, ensuring the best possible separation of samples \cite{bishop2006pattern, cortes1995support}.
    \item Random Forest is a learning algorithm that improves accuracy and manages overfitting by building multiple decision trees and identifying the most frequently occurring class among them. This method's ability to combine the predictions of many trees makes it a robust and reliable choice for classification tasks \cite{bishop2006pattern, breiman2001random}.
    \item Naive Bayes classifier is a probabilistic machine learning model based on Bayes' theorem. It calculates the probability of a class given a set of features, assuming that the features are independent. Despite this 'naive' assumption, the model often performs well in multiple real-world applications \cite{bishop2006pattern, hand2001idiot}.
\end{enumerate}

\begin{figure}[h!]
	\centering	\includegraphics[width=0.7\textwidth]{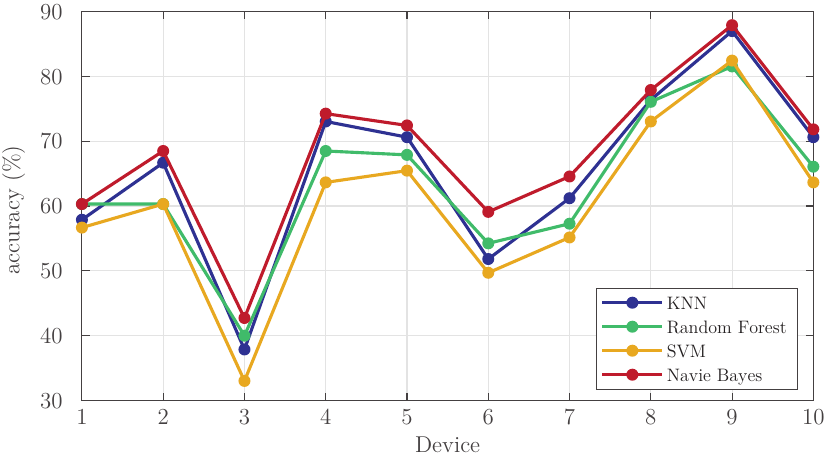}
\caption{Classification accuracy for the harvested energy datasets generated by the ten devices using: (a) KNN, (b) SVM, (c) Random Forest, and (d) Naive Bayes}
	\label{Ch5:fig:Class_Meth}
\end{figure}

Figure \ref{Ch5:fig:Class_Meth} displays the classification accuracy of the four methods applied to the energy data from the ten piezoelectric devices. Consistent with the previous discussion, the classification accuracy varies across datasets from different devices, largely due to overlapping energy values in certain fault condition labels. Moreover, the accuracy does not significantly differ between the classification methods, confirming that the energy values, as the unique feature, are the main factor influencing performance. In particular, Device 3 exhibits the lowest accuracy (from 33.03\% to 42.73\%), while Device 9 achieves the highest accuracy of 87.88\% using the Naive Bayes classifier. Since these two devices represent the extremes, we will focus on them to further investigate the factors behind the variability in the effectiveness of energy as a classification feature.

\begin{figure}[h]
	\centering	\includegraphics[width=1\textwidth]{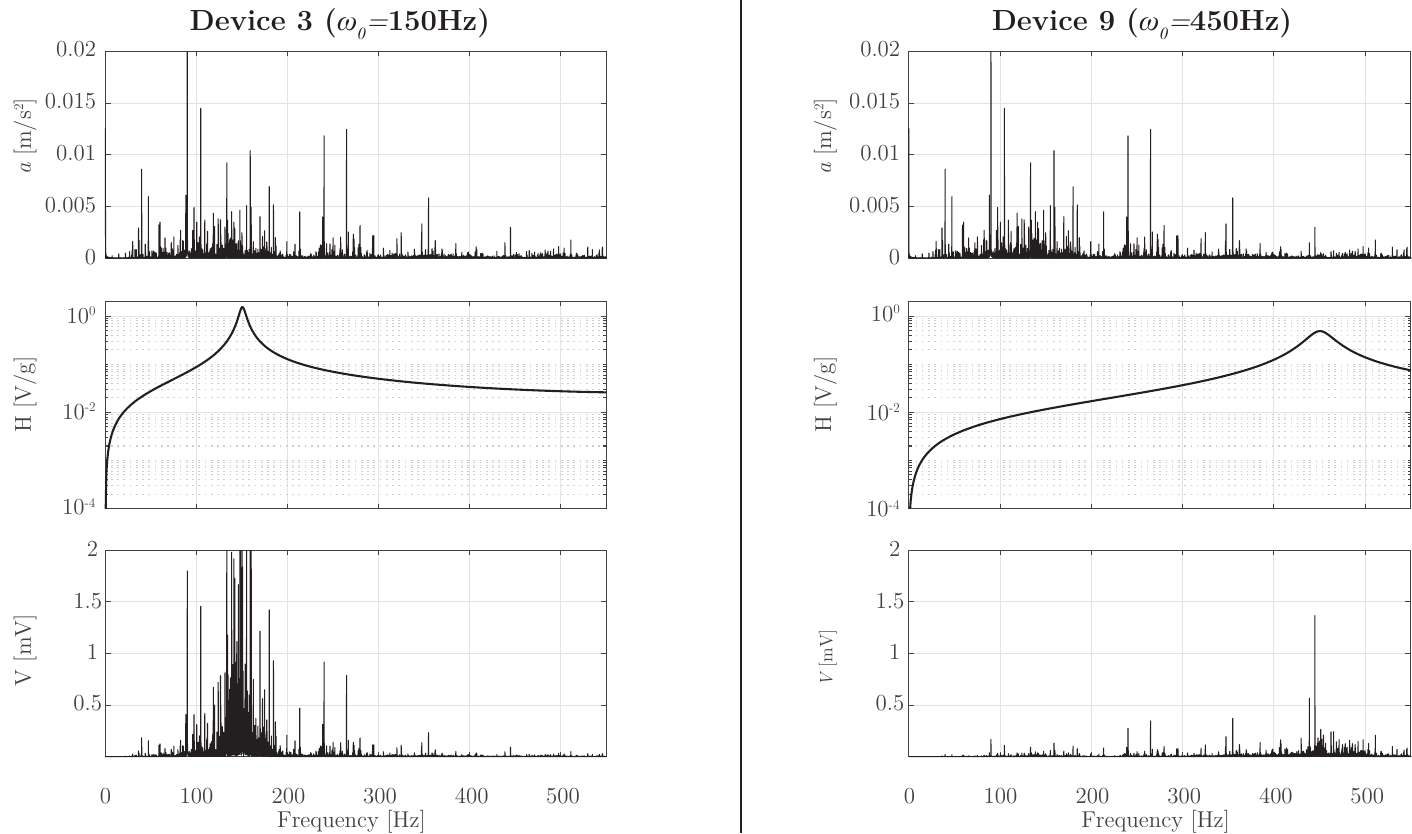}
\caption{Top: FFT of the 10-second signal in the normal condition (Label 7). Centre: FRF of Device 3 (left) and Device 9 (right). Bottom: FFT of the voltage generated by Device 3 (left) and Device 9 (right).}
	\label{Ch5:fig:accTOvol}
\end{figure}

The plots of the first row in Figure \ref{Ch5:fig:accTOvol} present the FFT of the 10-second signal under normal conditions (Label 7). Multiple harmonics are identified within the frequency range of 0 to 500 Hz, with a concentration around 100 Hz. The second row of Figure \ref{Ch5:fig:accTOvol} presents the FRFs of Devices 3 and 9, calculated using Eq. 3. The natural frequencies of the devices can be observed at frequencies where the voltage amplitude response relative to the acceleration amplitude response is maximised (50 Hz for Device 3 and 450 Hz for Device 9). The third row of Figure \ref{Ch5:fig:accTOvol} presents the FFT of the generated voltage, where only harmonics with significant amplitude around the device's natural frequency are observable. Specifically, the PEH transforms the acceleration signal into a voltage signal on a linear scale using the FRF. In this process, the PEH acts as a filter, amplifying the magnitude of harmonics near the natural frequency and suppressing those far from it, effectively preserving information within a specific frequency band.

\begin{figure}[h!]
	\centering	\includegraphics[width=1.0\textwidth]{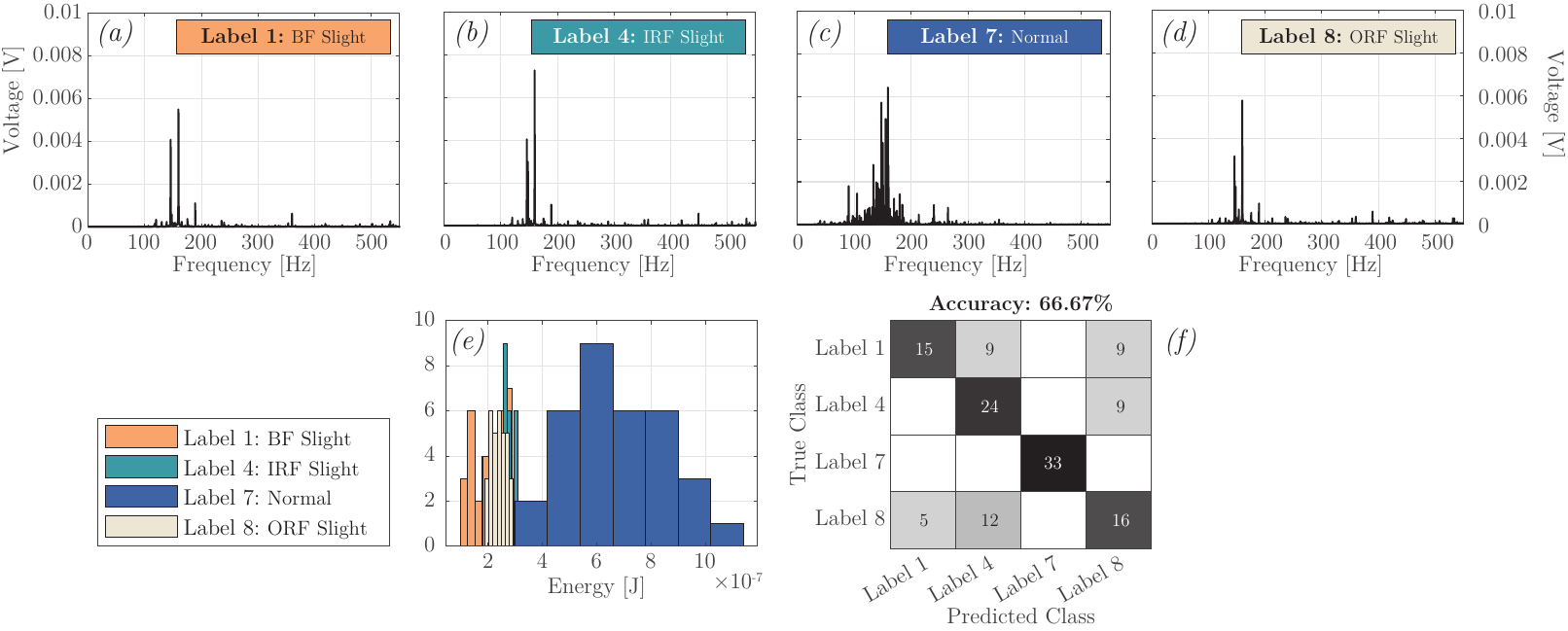}
\caption{FFT of the generated voltage for Device 3 under multiple conditions: $(a)$ Slight Ball Fault (Label 1), $(b)$ Slight Inner Race Fault (Label 4), $(c)$ Normal (Label 7), and $(d)$ Slight Outer Race Fault (Label 8). $(e)$ Histogram of the 33 events of 0.3 seconds for Labels 1, 4, 7, and 8. $(f)$ Confusion matrix for the Naive Bayes classifier implementation using K-fold with K = 5.}
	\label{Ch5:fig:D3}
\end{figure}

\begin{figure}[h!]
	\centering	\includegraphics[width=1.0\textwidth]{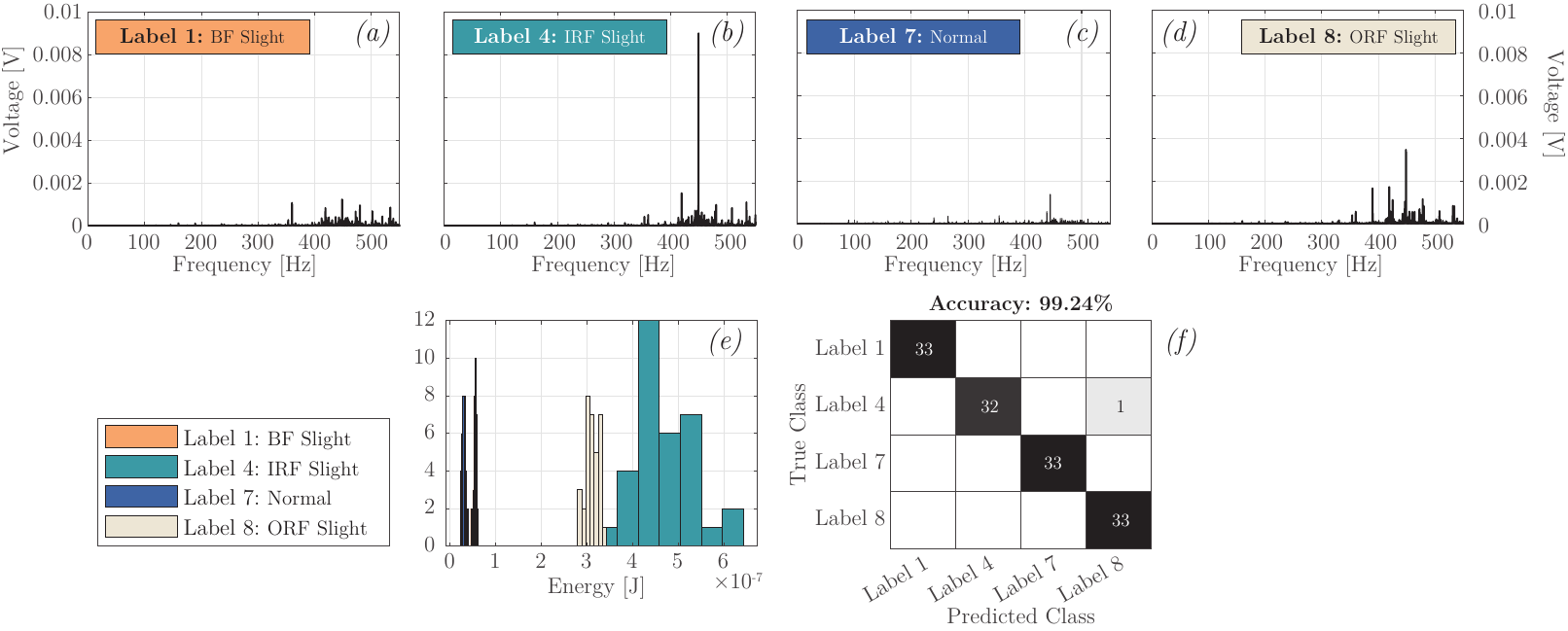}
\caption{FFT of the generated voltage for Device 9 under multiple conditions: $(a)$ Slight Ball Fault (Label 1), $(b)$ Slight Inner Race Fault (Label 4), $(c)$ Normal (Label 7), and $(d)$ Slight Outer Race Fault (Label 8). $(e)$ Histogram of the 33 events of 0.3 seconds for Labels 1, 4, 7, and 8. $(f)$ Confusion matrix for the Naive Bayes classifier implementation using K-fold with K = 5.}
	\label{Ch5:fig:D9}
\end{figure}

Figure \ref{Ch5:fig:D3}$a$–$d$ presents the voltage FFT for Device 3 under four conditions: normal, slight ball fault, slight inner race fault, and slight outer race fault. Only these four labelled states are shown to simplify the analysis, although the findings can be directly extrapolated to cases with ten labels. Analysing the FFT, the spectra corresponding to faulty bearing conditions (Figure \ref{Ch5:fig:D3}$a$–$c$) exhibit similar features, making it difficult to distinguish features for classification.  Moreover, Parseval's theorem allows us to directly estimate the harvested energy from the FFT as,
\begin{equation}
    E = c\int_{-\infty}^{\infty} \frac{V^2(\omega)}{R_l} d\omega \label{Ch5:eqEnergy2} 
\end{equation}
Thus, by analysing the FFT spectra, it is possible to identify which operational label generates more energy, with the normal condition (Label 7) producing the most. However, the energy output for the other three faulty labels is quite similar. Analysing histograms from 33-second datasets (Figure \ref{Ch5:fig:D3}$e$) confirms that Label 7 consistently produces more energy, with the energy from the remaining labels showing comparable magnitudes. This indicates that the spectral characteristics of the vibration signal around 150 Hz are not distinct enough to reliably differentiate between multiple states, which is reflected in the harvested energy. This is further supported by the results of the classification problem using Naive Bayes, as shown in Figure \ref{Ch5:fig:D3}$f$, where misclassifications occur primarily between the faulty labels.
Device 9 is subject to a similar analysis. Figure \ref{Ch5:fig:D9}$a$-$d$ presents the voltage FFT for the states labelled as Label 1, Label 4, Label 7, and Label 8 for Device 9. In this case, distinct characteristics in terms of amplitude and the presence of harmonics are easily identifiable. Moreover, since the harvested energy is proportional to the area under the squared spectrum curve, as per Parseval's theorem, the energy ranking from highest to lowest is: Label 8, Label 4, Label 1, and Label 7. This ranking is supported by Figure \ref{Ch5:fig:D9}$e$, which shows the histograms of events from the 33-second dataset. The energy ranking corresponds to the mean values of the histogram for the four classes. Additionally, the variability is not significant enough to cause overlap in the histogram, which positively impacts the classification accuracy. This is further confirmed in Figure \ref{Ch5:fig:D9}$f$, which presents the confusion matrix from the Naive Bayes method using K-fold cross-validation with K = 5, achieving an accuracy of 99.24\%.

Therefore, the analysis suggests that the primary reason for not achieving the perfect classification is the overlap in certain labelled conditions. Since the energy between different devices is not correlated, overlapping labels in some device datasets do not appear in others. This suggests that using multiple devices could enhance classification accuracy without the need to explore more complex classification methods; modify or add features; or simplify the classification problem by combining or reducing the labels of operational conditions.

\begin{figure}[h!]
	\centering	\includegraphics[width=0.9\textwidth]{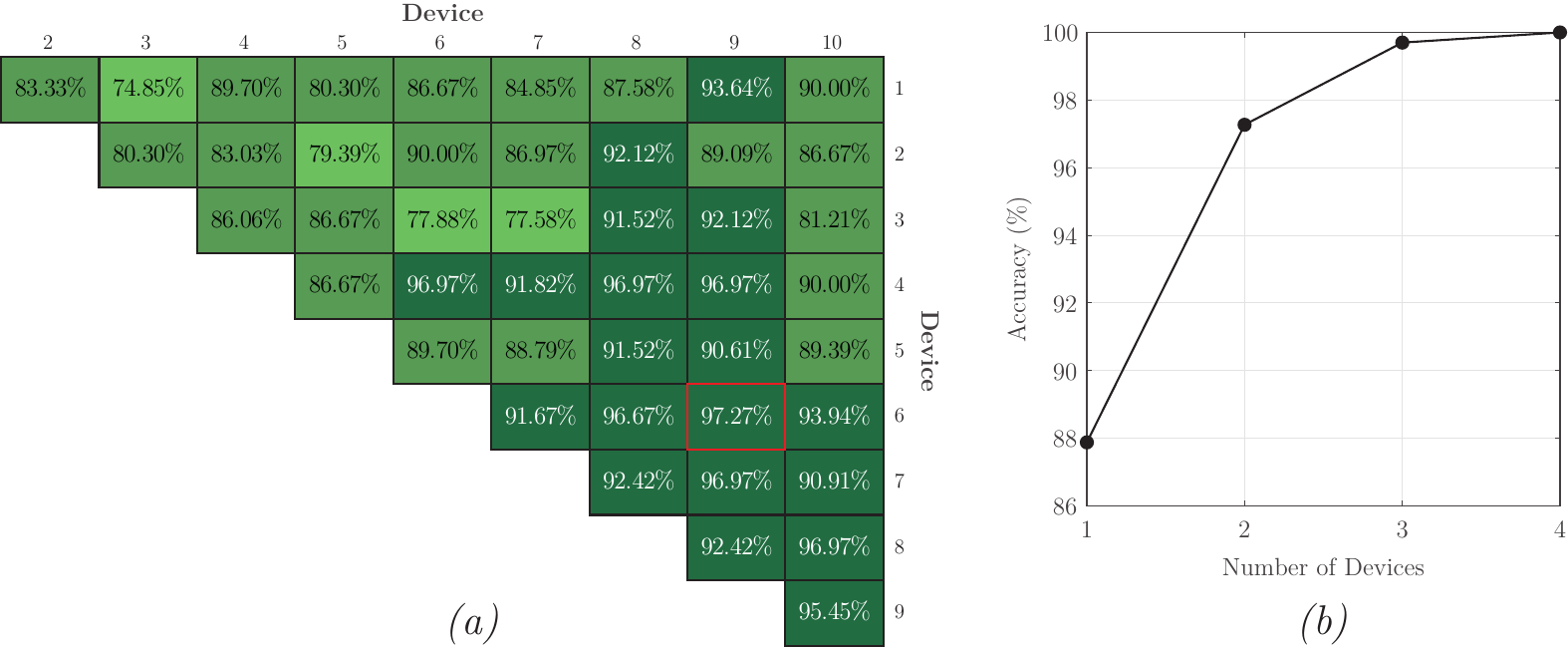}
\caption{$(a)$ K-fold validation accuracy for combinations of two piezoelectric devices $(b)$ The maximum accuracy achieved as a function of the number of devices used.}
	\label{Ch5:fig:mDevices}
\end{figure}

Figure \ref{Ch5:fig:mDevices}$a$ presents the accuracy of the Naive Bayes implementation using K-fold cross-validation (K = 5) for the combined harvested energy dataset from two devices for all possible unrepeated combinations and utilises the dataset with ten labels. The results indicate a consistent increase in accuracy when incorporating data from an additional device into the classification. Notably, the best combination is the set \{Device 9, Device 6\}, achieving an accuracy of 97.27\%. This supports the hypothesis that increasing the number of devices could potentially lead to an accuracy of 100\%. Figure \ref{Ch5:fig:mDevices}$b$ further explores this hypothesis and displays the accuracy of the optimal device combinations for the Naive Bayes implementation. Specifically, for three devices, the best sets are \{Device 4, Device 6, Device 8\} and \{Device 4, Device 6, Device 9\}, both reaching 99.7\%. For four devices, the best set is \{Device 4, Device 6, Device 8, Device 9\}, achieving 100\% accuracy. This validates the hypothesis and establishes a strong precedent for utilising harvested energy as a valuable feature for fault classification.

\section{Impact of the time window using augmented data for Bearing Fault Classification}
\label{Ch5:S:5}
An important factor to consider in this problem is the impact of the time window used for $event$ sampling from the database. Up to this point, we have focused on a time window of 0.3 seconds, constrained by the available data (10 seconds for the damage condition) and the number of samples needed to implement the classification algorithms. Since it is not feasible to directly increase the time window, we opted for a shorter time window of 0.1 seconds to get preliminary insights into the effect of this variable.

For each piezoelectric device, the new dataset contains harvested energy from 100 events, each lasting 0.1 seconds across the ten labelled conditions. To assess the performance in the classification task, the KNN algorithm is implemented, validating the results using 5-fold cross-validation. Table \ref{Ch5:t2:timewindows1} presents the classification accuracy and compares them with the results from the 0.3-second dataset in Section \ref{Ch5:S:4}. The results reveal that reducing the time windows leads to a drop in classification accuracy for all the devices. This outcome can be attributed to two main factors. Firstly, a longer time window allows small differences in the acceleration spectrum to be more clearly reflected in terms of energy, aiding in the differentiation of labels. Second, the harvested energy samples exhibit variability due to inconsistencies in the mechanism, which manifest in the acceleration response. Shortening the time window makes the magnitude of variability in the samples comparable to the mean value, thereby increasing the challenge of distinguishing different conditions. 

\begin{table}[h!]
\centering
\caption{Accuracy of the ten Devices using time windows of 0.1 and 0.3 seconds.}
\renewcommand{\arraystretch}{1.25}
\label{Ch5:t2:timewindows1}
\resizebox{\textwidth}{!}{
\begin{tabular}{rcrrrrrrrrr}
\hline
\multicolumn{1}{l}{}           & \multicolumn{10}{c}{Device}                                                                                                                                                                                                                           \\ \cline{2-11} 
\multicolumn{1}{c}{$\Delta t$} & 1                            & \multicolumn{1}{c}{2} & \multicolumn{1}{c}{3} & \multicolumn{1}{c}{4} & \multicolumn{1}{c}{5} & \multicolumn{1}{c}{6} & \multicolumn{1}{c}{7} & \multicolumn{1}{c}{8} & \multicolumn{1}{c}{9} & \multicolumn{1}{c}{10} \\ \hline
0.1 s                          & \multicolumn{1}{r}{42.90\%} & 54.70\%               & 26.80\%               & 63.20\%               & 51.00\%               & 40.50\%               & 35.90\%               & 58.80\%               & 69.20\%               & 60.20\%                \\
0.3 s                          & \multicolumn{1}{r}{57.88\%}  & 66.67\%               & 37.88\%               & 73.03\%               & 70.61\%               & 51.82\%               & 61.21\%               & 76.36\%               & 86.97\%              & 70.61\%                \\ \hline
\end{tabular}
}
\end{table}

\begin{figure}[h!]
    \centering	
    \includegraphics[width=1.0\textwidth]{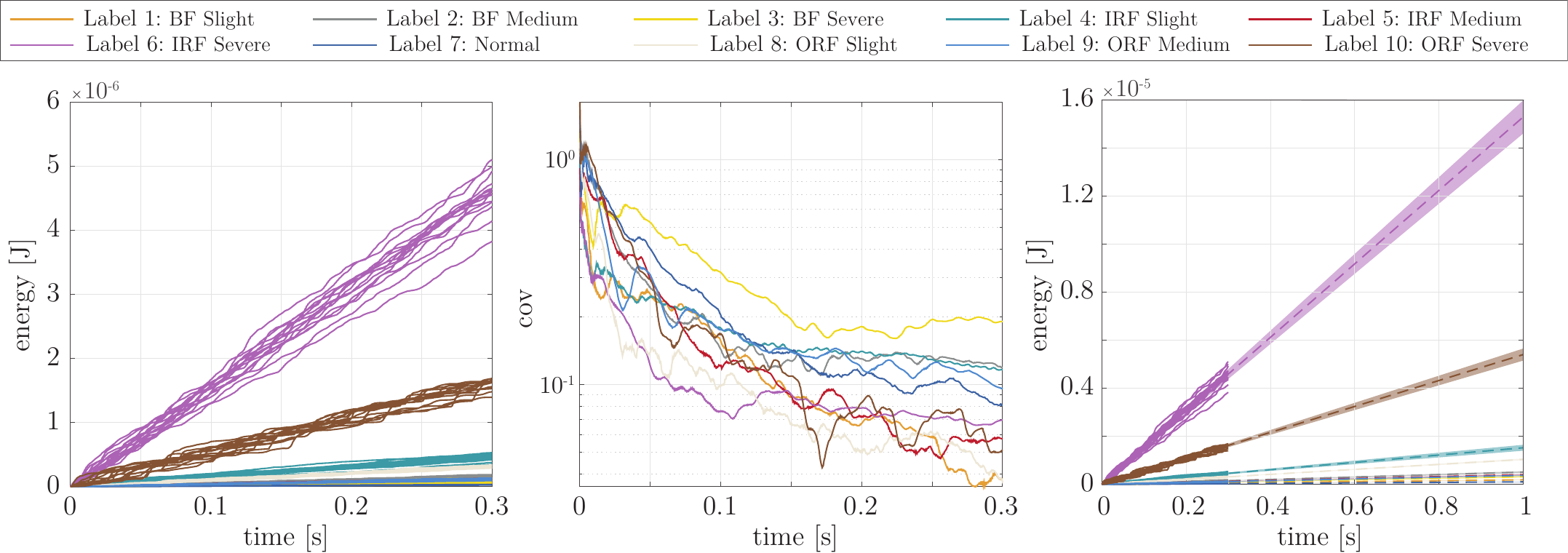}
    \caption{$(a)$ Accumulated harvested energy over time. $(b)$ Coefficient of variation of the accumulated energy over time. $(c)$ Projected expected accumulated energy and its confidence interval compared with the database.}
    \label{Ch5:fig:projected}
\end{figure}

Figure \ref{Ch5:fig:projected} supports these two statements by analysing Device 9, which has exhibited the best performance. Figure \ref{Ch5:fig:projected}$a$ presents the accumulated energy over time, revealing a linear increase. However, the slope—or rate of energy accumulation—varies depending on the fault condition. This aligns with the analysis in the previous section, which highlighted that the acceleration spectrum's characteristics and amplitude within the frequency range around the piezoelectric device’s resonance differ. As a result, the expected accumulated energy for each fault-bearing condition also follows a linear trend, with increasingly pronounced differences in amplitude magnitude over time. Moreover, Figure \ref{Ch5:fig:projected}$b$ presents the coefficient of variation (CoV) of the accumulated energy over time for various bearing conditions. The CoV values decrease over time, confirming the second statement. This trend aids in distinguishing between labels, as the probability of value overlap decreases with an extended time window. Given the linear increase and decreasing variability, the expected accumulated energy and its confidence interval can be projected. Figure \ref{Ch5:fig:projected}$c$ illustrates this, with the projected accumulated energy represented by a dotted line and the 98\% confidence interval shown as a shaded area.  Additionally, Table \ref{Ch5:t3:slopes} presents the slope of the estimated expected accumulated energy, where it is possible to identify the dissimilarity between them. In this sense, extending the time window to 1 second is expected to improve accuracy, approaching 100\% significantly.

\begin{table}[h!]
\centering
\caption{Slopes of the accumulated energy over time for Device 9 under each labeled fault condition}
\renewcommand{\arraystretch}{1.25}
\label{Ch5:t3:slopes}
\begin{tabular}{lllllllllll}
\hline
Device                            & \multicolumn{1}{c}{1} & \multicolumn{1}{c}{2} & \multicolumn{1}{c}{3} & \multicolumn{1}{c}{4} & \multicolumn{1}{c}{5} & \multicolumn{1}{c}{6} & \multicolumn{1}{c}{7} & \multicolumn{1}{c}{8} & \multicolumn{1}{c}{9} & \multicolumn{1}{c}{10} \\ \hline
Slope {[}$\mu$J/s{]} & 0.19                  & 0.51                  & 0.32                  & 1.52                  & 0.40                  & 15.30                 & 0.10                  & 1.04                  & 0.37                  & 5.40                   \\ \hline
\end{tabular}
\end{table}

Based on these findings, it is interesting to explore longer time windows, as we expect classification accuracy to improve in correlation with the extended time window. To overcome data limitations, we implemented the StefielGen data augmentation method \cite{cheema2024stiefelgen} to generate synthetic data and extend our study to longer time windows.

Data augmentation based on StefielGen algorithm is a novel approach presented in \cite{cheema2024stiefelgen}. The method stands out due to its simplicity, flexibility, lack of a need for model training, and interpretability. The process consists of three key steps: (1) Time Series Decomposition, (2) Matrix Perturbation, and (3) Augmented Time Series Reconstruction. Further explanations are available in Appendix A.

In particular, $\beta$ is a hyperparameter in the augmentation algorithm that plays a crucial role in generating signals with varying degrees of novelty. According to \cite{cheema2024stiefelgen}, higher $\beta$ values are preferable for creating signals that require a high degree of novelty, while smaller values are more suitable for standard data augmentation. To evaluate the impact of $\beta$, three values—0.1, 0.4, and 0.7—are selected, corresponding to low, moderate, and large perturbations, respectively. We start with a preliminary study which aims to replicate the 10-labels classification results using the 0.3-second dataset from Section \ref{Ch5:S:4} with augmented data to ensure reproducibility. Specifically, ten 1-second acceleration signals are sampled from the original database. These samples are randomly split into a seed database and a testing database. The seed database is used to create an augmented database, consisting of new synthetic acceleration signals, to replicate the study in Section 4. Note that another important hyperparameter to define in the augmentation method is the dimensionality of the reshaped matrices. In this implementation, the chosen dimensions are 150$\times$100. Additionally, from the testing data, 150 events (15 per label), each lasting 0.3 seconds, are sampled in accordance with the dataset’s time-length constraints to test the classification model (trained with the augmented data) using data that was not part of the creation of the synthetic signals. The analysis focuses on Device 9  in the previous section, which demonstrated the best performance, allowing its results to be extrapolated to other devices. The studies were conducted using a Naive Bayes classifier and validation accuracy is determined using 5-fold cross-validation, while testing accuracy is computed using the testing dataset on a model trained with the entire augmented dataset. This approach provides a comprehensive evaluation of the model’s performance and assesses the effectiveness of the augmentation algorithm across various perturbation levels.

\begin{figure}[h!]
    \centering	
    \includegraphics[width=0.7\textwidth]{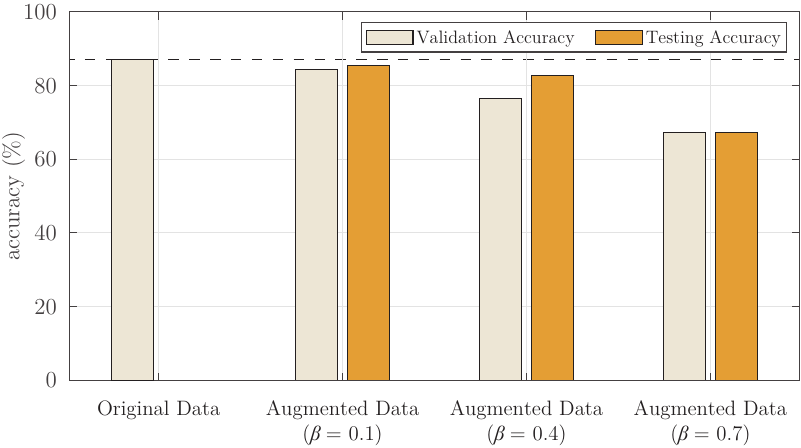}
    \caption{Validation accuracy of the original dataset of 0.3 seconds events, compared with the validation and testing accuracy of augmented data across varying levels of perturbation ($\beta$: 0.1, 0.4, and 0.7) in Device 9.}
    \label{Ch5:fig:beta_study}
\end{figure}

Figure \ref{Ch5:fig:beta_study} presents the validation accuracy for both the original and augmented data, along with the testing accuracy for the testing data. The dotted line indicates the validation accuracy of the original data. The results reveal that low beta values produce accuracy comparable to the original data while increasing beta leads to a decline in accuracy. These findings lead us to select a $\beta$ value of 0.1 for further analysis. Furthermore, there is a higher correlation between the validation accuracy using the augmented data and the testing accuracy using data from the original dataset that did not participate in the augmentation process.

Therefore, after explaining the augmentation method and selecting the hyper-parameters, we can continue the discussion about the impact of time windows. Augmented data from the seed dataset is used to tackle the classification problem across various time windows. Table \ref{Ch5:t:numberSample} lists the sizes of the augmented data for training and validation, as well as the testing data for the different time windows. It is important to note that the number of events in the training data is constrained by the fact that the testing data consists of five 1-second signals.

\begin{table}[h!]
\centering
\caption{Number of sample events for different time windows.}
\renewcommand{\arraystretch}{1.25}
\label{Ch5:t:numberSample}
\begin{tabular}{lrrrr}
\hline
               & \multicolumn{1}{l}{0.1 seconds} & \multicolumn{1}{l}{0.3 seconds} & \multicolumn{1}{l}{0.5 seconds} & \multicolumn{1}{l}{1 second} \\ \hline
Original Data  & 1000                            & 330                             &   -                          & -                         \\
Augmented Data & 1000                            & 330                             & 300                             & 300                          \\
Testing Data   & 500                             & 150                             & 100                             & 50                           \\ \hline
\end{tabular}
\end{table}

\begin{figure}[h!]
    \centering	
    \includegraphics[width=1.0\textwidth]{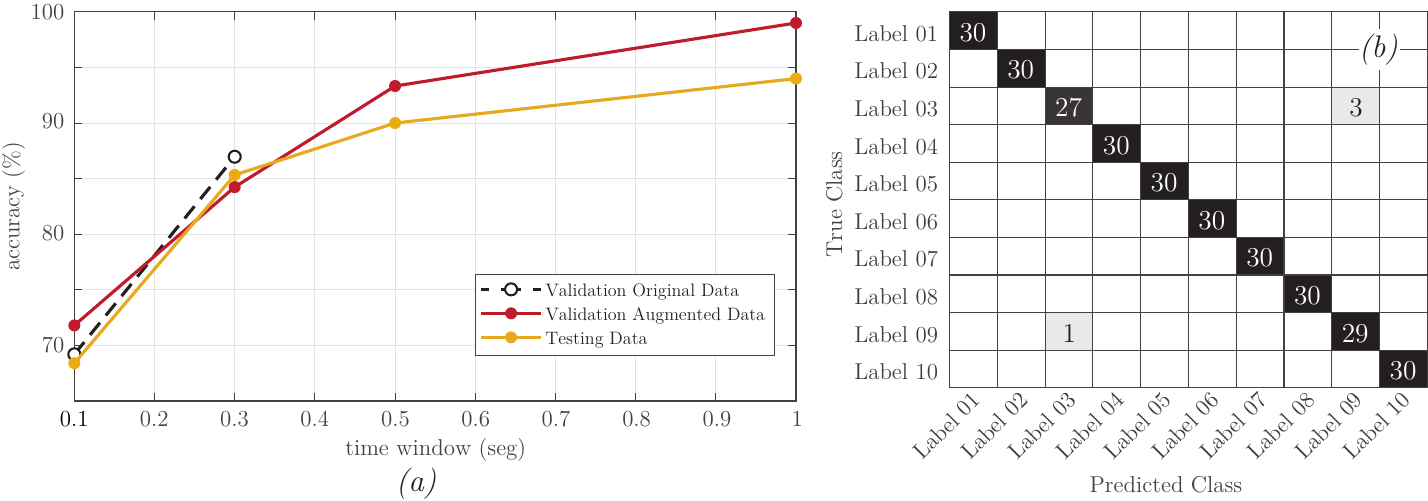}
    \caption{$(a)$ Validation and testing accuracy of augmented datasets with different time windows in Device 9. $(b)$ Confusion Matrix from the Validation Database Using 1-Second Time Windows.}
    \label{Ch5:fig:t_vs_acc}
\end{figure}

Figure \ref{Ch5:fig:t_vs_acc} presents the validation accuracy for both the original and augmented data using K-fold cross-validation (K = 5), along with the testing accuracy for the entire augmented dataset. Complementing the findings from the $\beta$ study, the validation accuracy of the original data aligns with the validation and testing results from the training data using augmented data for time windows of 0.1 and 0.3 seconds. Additionally, as expected, the accuracy of the augmented data increases with larger time windows. However, the testing accuracy shows less consistency with the validation accuracy as the time window increases. This discrepancy arises because the number of events in the testing dataset decreases with larger time windows due to data limitations, making the testing results less representative of the classification performance.


\section{Utilising Capacitor Energy for Bearing Fault Classification}
\label{Ch5:S:6}
In prior sections, the analysis has focused on the energy of the voltage generated by a piezoelectric device directly connected to a load resistor ($R_l$). However, this approach presents two notable limitations. First, energy harvesting applications generally require the rectification and storage of the generated voltage to utilise the energy in transmitters \cite{covaci2020piezoelectric}. Second, this study aims to estimate the energy feature using a hardware component instead of post-processing the instantaneous voltage to minimise the transmitted data. Despite these limitations, the current analysis is insightful, as it allows for the derivation of general guidelines on the performance of piezoelectric devices in classification tasks related to bearing fault detection, independent of specific electrical circuits.

\begin{figure}[h]
    \centering	
    \includegraphics[width=0.7\textwidth]{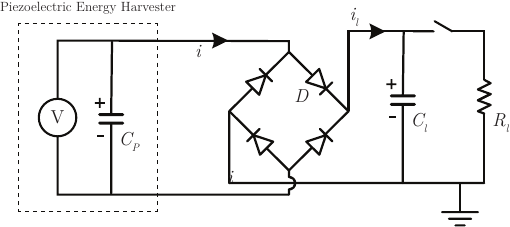}
    \caption{Schematic diagram of a standard energy harvesting circuit.}
    \label{Ch5:fig:SEHcircuit}
\end{figure}

Therefore, this section evaluates the guidelines previously established by testing them with the energy accumulated in a capacitor within an electrical circuit designed for energy-harvesting applications. The selected configuration is the Standard Energy Harvesting (SEH) circuit, one of the most commonly adopted designs, illustrated in Figure \ref{Ch5:fig:SEHcircuit}. The SEH circuit comprises a full-bridge rectifier, a storage capacitor ($C_l$), an electrical resistor ($R_l$), and a switch that closes once the capacitor achieves the desired charge level. In particular, in this study, the transient energy accumulated in the capacitor is utilized for classification purposes rather than relying on the steady-state power across the load resistor.

In this case, the electrical equation, including the mechanical coupling as presented in Eq. \ref{Ch5:SistEq2Rnew}, is generalised, as the electrical current in the piezoelectric system can no longer be expressed simply as $v(t)/R_l$ :
\begin{equation}
\label{Ch5:SistEq2Rnew}
    C_p\dot{v}(t) + \boldsymbol{\Theta}^T\boldsymbol{\Phi}_o \dot{\boldsymbol{\eta}}= - i(t)
\end{equation}

The current $i_l(t)$ through the capacitor depends on the relative magnitudes of the piezoelectric voltage $v(t)$, the capacitor voltage $v_c(t)$, and the diode's forward voltage drop $v_{fd}$. When $v(t) > v_c(t) + v_{fd}$, the rectifier diodes conduct, allowing current to flow into the capacitor. Conversely, when $v(t) \leq v_c(t) + v_{fd}$, the diodes block the current, and $i_l(t) = 0$. This behaviour is described by the following equation:

\begin{equation}
i_l =
\begin{cases}
\frac{v - v_{fd} - v_c}{R_{on}}, & \text{if } v > v_c + v_{fd}, \\
0, & \text{if } V \leq v_c + v_{fd},
\end{cases}
\end{equation}
where $R_{on}$ is the diode's on-state resistance. This current charges the capacitor, and the resulting capacitor voltage \( v_c \) evolves according to the relationship:
\begin{equation}
C \frac{dv_c}{dt} = i_l
\end{equation}
where $C$ is the capacitance. This dynamic reflects the nonlinear nature of the diode rectification and the energy transfer from the piezoelectric element to the storage capacitor.

Clementino et al. \cite{clementino2014modeling} presented a Simulink block diagram to estimate the voltage accumulated in the capacitor by solving a state-space formulation (Eq. \ref{Ch5:eq:state-space}) using MATLAB's \verb!ode45! solver. The state-space formulation integrates the dynamic equation with electrical coupling and incorporates the redefined electrical circuit equation with mechanical coupling, providing a comprehensive representation of the system dynamics, as follows:
\begin{equation}
        \dot{\textbf{x}}  = \boldsymbol{A}\textbf{x} + \boldsymbol{B}\textbf{u}
        \label{Ch5:eq:state-space}
\end{equation}
where $\textbf{x} = \{\boldsymbol{\eta}(t)\quad \boldsymbol{\dot{\eta}}(t)\quad v(t)\}$ is the state-space vector,  $\textbf{u} = \{a(t)\quad i(t)\}$ is the input vector and the matrices $\boldsymbol{A}$ and $\boldsymbol{B}$ are defined as follows,
\begin{equation}
    \boldsymbol{A} = 
    \begin{bmatrix}
    \textbf{0}_{K\times K} & \textbf{I}_{K\times K} & \textbf{0}_{K\times 1}\\
    -\textbf{k}_o & -\textbf{c}_o & \boldsymbol{\theta}_o\\ 
    \textbf{0}_{1\times K} & -\frac{\boldsymbol{\Theta}^T\boldsymbol{\Phi}_o}{C_p}& 0
    \end{bmatrix}, \quad
    \boldsymbol{B} = 
    \begin{bmatrix}
    \textbf{0}_{K\times 1} & \textbf{0}_{K\times 1} \\
    \textbf{f}_o & \textbf{0}_{K\times 1} \\ 
    0 & C_p^{-1}
    \end{bmatrix}
\end{equation}


After presenting the methodology for estimating capacitor voltage, a series of studies were conducted to demonstrate the performance of this specific SEH circuit with a capacitance of 10 $\mu$F. The acceleration database used in this study consists of augmented data from 1-second time windows, as employed in the previous section. The studies were conducted using a Naive Bayes classifier and validated through 5-fold cross-validation. Our focus is on Device 9, which exhibited the best performance in previous studies. Similar to the previous case, our classification feature is the cumulative energy, this time measured in the capacitor. This measurement is effectively equivalent to using the voltage, as expressed in the following equation, 
\begin{equation}
    E = \frac{1}{2}Cv_c^2
\end{equation}

\begin{figure}[h]
    \centering	
    \includegraphics[width=1.0\textwidth]{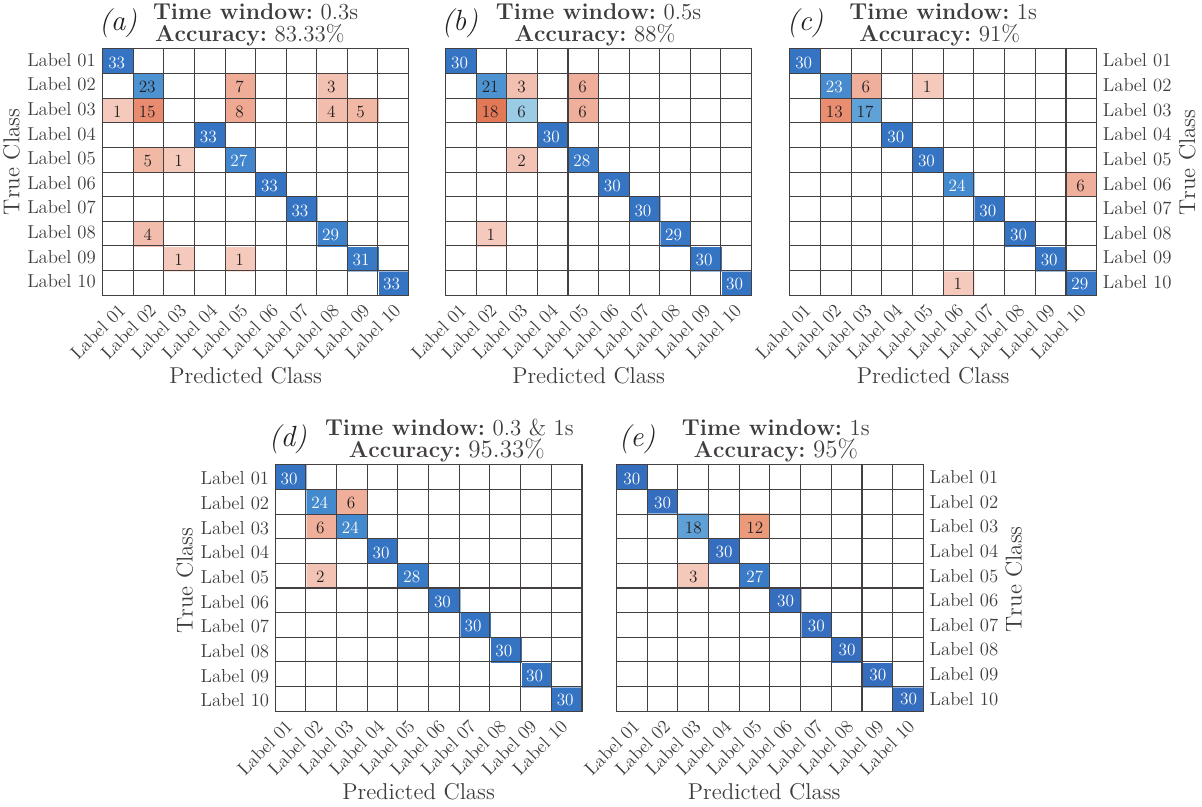}
    \caption{Confusion matrices and validation accuracies for Device 9, using capacitor of 10 $\mu$F energy from: $(a)$ the original dataset with a 0.3-second time window $(b)$ the augmented dataset with a 0.5-second time window $(c)$ the augmented dataset with a 1-second time window $(d)$ the augmented 1-second time window dataset, incorporating capacitor energy at both 0.3 and 1 second. $(e)$ Confusion matrices and validation accuracies for Device 9, using capacitor of 1000 $\mu$F energy from the augmented dataset with a 1-second time window.}
    \label{Ch5:fig:confusion_capa}
\end{figure}

\begin{figure}[h!]
    \centering	
    \includegraphics[width=1.0\textwidth]{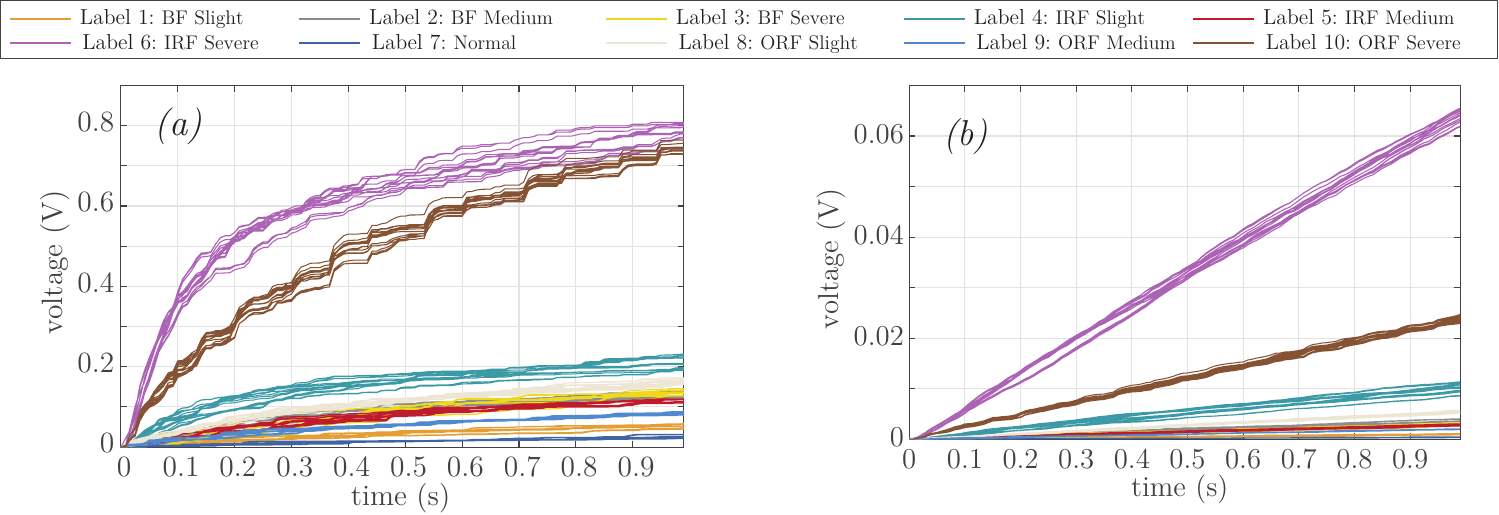}
    \caption{Capacitor voltage over time for 300 sample events within a 1-second augmented dataset, for capacitors with capacitances of: $(a)$ 10$\mu$F and $(b)$ 1000$\mu$F.}
    \label{Ch5:fig:capacitor_voltage}
\end{figure}

Figure \ref{Ch5:fig:confusion_capa} shows confusion matrices for four different time windows. Figure \ref{Ch5:fig:confusion_capa}$a$ presents the confusion matrix for the original dataset with a 0.3-second time window. The accuracy achieved is 83.33\%, a decrease compared to the implementation of Section 4 (87.88\%). The primary issue lies in the classification of Label 3, which represents severe ball damage. Increasing the time window can help address this problem. Figure \ref{Ch5:fig:confusion_capa}$b$ shows the confusion matrix for the augmented dataset with a 0.5-second time window. As expected, the accuracy improves to 88\%, and the classification issues of Label 3 are mitigated. Figure \ref{Ch5:fig:confusion_capa}$c$ presents the confusion matrix for the augmented dataset with a 1-second time window. While the accuracy further increases to 91\%, new misclassifications occur between Labels 6 and 10. To understand this issue, Figure \ref{Ch5:fig:capacitor_voltage}$a$ displays the instantaneous capacitor voltage for the first 300 samples of the 1-second dataset. It is possible to observe that Label 10 events rapidly reach a voltage level, while Label 6 events require approximately 0.6 seconds to reach a similar voltage level. This explains the misclassification between these labels at specific time window values. This suggests that considering the capacitor energy at multiple time points, such as 0.3 and 1 second, can improve classification accuracy. Figure 20d presents the confusion matrix for the augmented 1-second dataset, using the harvested energy at both 0.3 and 1 second. The accuracy exceeds previous results, reaching 95.33\%, confirming our observation. A different approach to deal with this issue is ensuring the capacitor voltage's linearity, as Lan et al. \cite{lan2017capsense} elaborated. In particular, the capacitor voltage $V_c(t)$ during charging follows an exponential behaviour, given by
\begin{equation}
    v_c(t) = v_{max}(1-e^{-t/\tau})
\end{equation}
here, $v_{max}$ is the maximum voltage to which the capacitor can be charged, defined as $v_{max} = \min\{v_s, v_r\}$, where $v_s$ is the applied voltage (in this case, the rectified piezoelectric voltage), and $v_r$ is the capacitor's rated voltage. Additionally, $\tau$ is the \textit{time constant}, given by $\tau = RC$, where $R$ is the equivalent resistance of the resistor-capacitor charging circuit, and $C$ is the capacitance of the capacitor. In particular, during the charging process, the voltage follows a behaviour that can be approximated as linear between the times 0 and $\tau/2$. To extend this linear behaviour and avoid overlapping after a short time, one option is to increase the capacitance value based on the definition of $\tau$. For instance, consider a capacitance of 1000 $\mu F$. Figure \ref{Ch5:fig:confusion_capa}$e$ presents the confusion matrix of the augmented data from 1-second time windows using a capacitance of 1000 $\mu F$. It is possible to identify that the accuracy reaches 95\%, mainly due to ensuring the linearity of the capacitor voltage. This is confirmed in Figure \ref{Ch5:fig:capacitor_voltage}$b$, which shows the voltage signal of the 300 events over time, confirming the linear behaviour

\begin{figure}[h!]
    \centering	
    \includegraphics[width=0.7\textwidth]{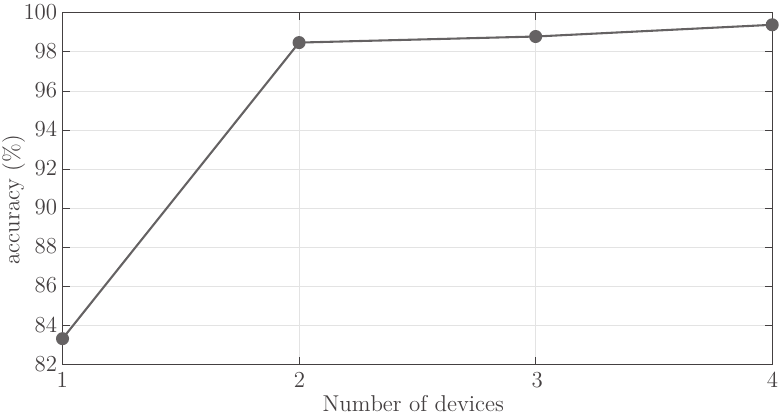}
    \caption{Validation accuracy of the original 0.3-second time window datasets considering different number of devices}
    \label{Ch5:fig:Ndev_cap}
\end{figure}

Another important conclusion from previous studies is that increasing the number of devices can improve classification accuracy. To test these conclusions, the device sets identified in Section 4 are considered: \{Device 6, Device 9\}, \{Device 4, Device 6, Device 9\}, and \{Device 4, Device 6, Device 8, Device 9\}. Figure \ref{Ch5:fig:Ndev_cap} presents the accuracy for these device sets. As observed, the accuracy increases significantly from the single-device case to 98.48\% with two devices. However, further increasing the number of devices to four only yields a marginal improvement, reaching 98.79\% and 99.39\%, respectively. This confirms the conclusion that increasing the number of devices can enhance classification performance, although perfect classification may not always be achievable with four devices.
.

\section{Anomaly detection based on the Harvested Energy}
\label{Ch5:S:7}
In certain contexts, anomaly detection models are preferred because it is often impossible to anticipate all fault scenarios fully or because collecting data for every possible scenario can be challenging. These models offer significant flexibility by enabling alerts to changes in conditions, even those not previously anticipated. In Section \ref{Ch5:S:4}, the analysis reveals that specific piezoelectric devices can effectively distinguish between labelled conditions. Specifically, the energy output clearly distinguishes between the healthy state and damaged scenarios. This distinction is evident in the histograms for Devices 1 and 2 shown in Figure \ref{Ch5:fig:EnergyHist}, where a limit can be defined to separate healthy events from damaged ones. However, the identification and definition of this limit must be conducted systematically and rigorously. This section presents an implementation of anomaly detection based on harvested energy.

The studies in Section \ref{Ch5:S:4} reveal that the Naive Bayes classifier method outperforms other algorithms. Leveraging the core concept of this method, we can define a straightforward algorithm to tackle the anomaly detection problem. This algorithm involves fitting a Gaussian probabilistic function to the harvested energy data for the bearing’s normal operating state. By using the Z-score normalisation \cite{montgomery2010applied}, which quantifies the distance between a harvested energy sample ($E_i$) and the mean ($\mu$) in units of standard deviation ($\sigma$), we can determine if a harvested energy sample significantly deviates from the mean, thus identifying it as an anomaly.
\begin{equation}
    Zscore = \frac{E_i-\mu}{\sigma}
\end{equation}

In this study, a Z-score threshold of 3 is employed. Events with harvested energy within this threshold are classified as healthy. Conversely, events exceeding this threshold are categorised as damaged.

\begin{figure}[h]
	\centering
	\includegraphics[width=0.9\textwidth]{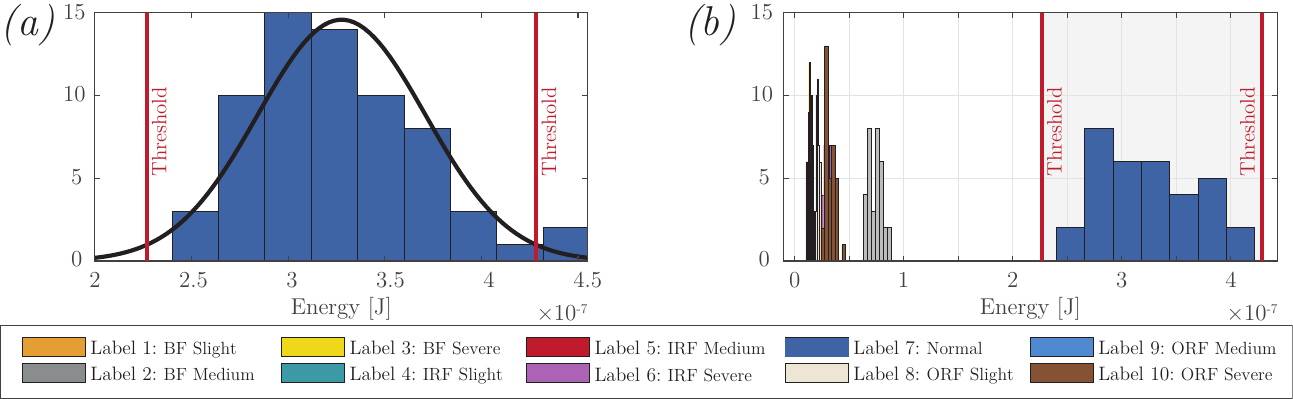}
\caption{Gaussian anomaly detection results. Histogram of the 30 events from the healthy dataset, showing their fitted normal distribution and the threshold set at the 98\% confidence level (right). Histogram of the 10 datasets corresponding to the state condition labels, with the threshold defined using Gaussian anomaly detection (left)}
	\label{Ch5:fig:Anomaly}
\end{figure}

As previously discussed, Device 3 is an ideal candidate for implementing the anomaly detection algorithm because the histograms of its normal operating conditions are clearly distinguishable from those of other cases. On the other hand, as mentioned in Section \ref{Ch5:S:3}, 40 seconds of acceleration response data under normal conditions are available; however, only 10 seconds were used to ensure a balanced dataset. Despite this, the remaining 30 seconds are utilized for an example implementation when defining the threshold. In this implementation, 100 events were sampled, the normal distribution of the dataset was estimated, and a threshold was defined with Z-score of 3, as illustrated in Figure \ref{Ch5:fig:Anomaly}$a$. This defined range was then applied to the dataset from the previous section for classification, resulting in 100\% accuracy in anomaly detection, as shown in Figure \ref{Ch5:fig:Anomaly}$b$.

\section{Conclusions}
\label{Ch5:S:8}
This study demonstrated the feasibility of using harvested energy from piezoelectric harvesters (PEHs) for fault detection. By integrating energy harvesting with fault detection, the proposed approach addresses critical challenges in the reliability of rotating machinery, aiming to develop self-powered sensors. The study evaluated the performance of various PEH designs with different natural frequencies, utilizing a numerical model of PEHs and acceleration data from the CWRU dataset.

First, the harvested energy was tested as a classification feature using different algorithms. The one with the best performance was the Naive Bayes methodology, achieving an accuracy of 87\%. Subsequently, the incorporation of multiple devices was studied, concluding that the accuracy increased to 100\% when four devices were used.

Moreover, by effectively implementing a methodology to augment the CWRU dataset, the time windows used to calculate the harvested energy were studied. The study revealed that increasing the time windows has a positive impact, enhancing the overall accuracy. For a single device, it is possible to increase the accuracy to approximately 99\% increasing the time windows. 

The obtained conclusions are used to analyse capacitor energy as a classification feature, demonstrating the adaptability of PEH designs in energy harvesting circuits for practical applications. This is particularly valuable because the harvested energy is calculated analogously using passive electrical components, which significantly reduces the need for data digitalization and transmission, thereby saving energy. The framework successfully reduced the data required for decision-making from 20 million bytes to just 2 bytes.


\newpage
\appendix
\section{StefielGen algorithm}

\begin{enumerate}
    \item \textbf{Time Series Decomposition}: The time series vector $S_1 \in \mathbb{R}^{p \times 1}$ is reshaped into a matrix $T_1 \in \mathbb{R}^{m \times n}$, where $m$ and $n$ are hyperparameters of the method. If the length of the time series vector $p$ is not an exact multiplier of $m \cdot n$, two strategies can be employed: $(i)$ padding the sequence, or $(ii)$ introducing a slight overlap in the signal. The resulting matrix $T_1$ is then decomposed using Singular Value Decomposition (SVD) into $T_1 = U_1 \Sigma V_1^T$, where $U_1 \in \mathbb{R}^{m \times n}$ and $V_1 \in \mathbb{R}^{m \times n}$ belong to the Stiefel manifold ($St_n^m$), which is the set of all orthogonal matrices of size $m \times n$, i.e., 
    \begin{equation}
        St_n^m := \{U\in \mathbb{R}^{m \times n}\;|\; U^TU = I_n, m>n\}
    \end{equation}
    Meanwhile, $\Sigma \in \mathbb{R}^{m \times n}$ is a diagonal matrix containing the singular values, representing the magnitude of the data along each principal direction.
    \item \textbf{Matrix Perturbation}: This step leverages the fact that $U_1$ and $V_1$ are Stiefel matrices. Let's focus on $U_1$ and consider a randomly sampled matrix $\Delta_1\in \mathbb{R}^{m \times n}$ from the Stiefel tangent space at $U_1$, $\mathcal{T}_{U_1} St_n^m$, defined as:
    \begin{equation}
        \mathcal{T}_U St_n^m := \{\Delta\in \mathbb{R}^{m \times n}\;|\; \Delta^TU + U^T\Delta = 0\}
    \end{equation}
    In the Stiefel manifolds, it is possible to define a geodesic, which corresponds to the length-minimizing curve within the manifold, and whose shape is governed by a Riemannian metric, reflecting the manifold's curvature. Geodesics gives rise to the Riemannian exponential at a base $U$, $\text{Exp}_{U}(\Delta)$,
    \begin{equation}
         \Tilde{U} = \text{Exp}_U(\Delta) = (U\quad \Delta)\ \text{exp}_m 
         \begin{pmatrix}
            U^T\Delta & -\Delta^T\Delta \\
            I_m & U^T\Delta
        \end{pmatrix}
        \begin{pmatrix}
            I_m \\
            0 
        \end{pmatrix}
        \text{exp}_m(-U^T\Delta)
    \end{equation}
    The Riemannian exponential map at a base point $U$ moves in the direction of the tangent matrix $\Delta$ within the Stiefel manifold, mapping to an endpoint matrix $\Tilde{U} \in \mathbb{R}^{m \times n}$. 
    Specifically, the radius of injectivity defines the infimum radius around $U$ within which the Riemannian exponential map \(\text{Exp}_U\) remains a diffeomorphism with the Stiefel manifold $St_n^m$. This radius has a global value of \(0.89\pi\). Consequently, the tangent matrix $\Delta_1$ is normalized with respect to the canonical metric, yielding the matrix $\Bar{\Delta}$, as:
        \begin{equation}
        \Bar{\Delta} = \frac{\Delta}{ \sqrt{\langle \Delta, \Delta \rangle_U} } = \frac{\Delta}{|| \Delta ||_U}
    \end{equation}
    where,
    \begin{equation}
        \langle \Delta, \Delta \rangle_U = \text{tr}\left(\Delta^T\left(I-\frac{1}{2}UU^T\right)\Delta\right)
    \end{equation}
    Therefore, the perturbed matrix $U_2\in \mathbb{R}^{m \times n}$ is then calculated as follows:
    \begin{equation}
         U_2 = \text{Exp}_{U_1}(0.89\pi\beta\Bar{\Delta}) 
    \end{equation}
    Note that $U_2$ belongs to the Stiefel manifold, and $\beta$ is a hyperparameter of the method. An analogous procedure is used to calculate the perturbed matrix $V_2$ from $V_1$.
    \item \textbf{Augmented Time Series Reconstruction}: The final step involves reconstructing the augmented time series vector. The new matrices $U_2$ and $V_2$ are used along with the original matrix $\Sigma$ from the SVD of matrix $M_1$. The new matrix $M_2\in \mathbb{R}^{m \times n}$ is defined as:
    \begin{equation}
        M_2 = U_2 \Sigma V_2^T
    \end{equation}
    $M_2$ is then reshaped to obtain the augmented time series vector $S_2 \in \mathbb{R}^{p \times 1}$, completing the process of generating synthetic data.
\end{enumerate}

\bibliographystyle{model1-num-names}
\bibliography{reference.bib}

\end{document}